\documentclass[12pt,a4paper]{article}
\usepackage[utf8]{inputenc}
\usepackage[total={6in, 7.9in}]{geometry}
\title{\textbf{The Aharonov-Bohm effect \\in a closed flux line}}
\author{Ricardo Heras\thanks{e-mail: ricardo.heras@ou.ac.uk}}
\date{{\small \textit{School of Physical Sciences, The Open University,\\ Walton Hall, Milton Keynes MK7 6AA, UK}}}

\usepackage{bm}
\usepackage{graphicx}
\usepackage[mathscr]{euscript}
\usepackage{amsmath}
\usepackage{enumitem}
\usepackage{amsfonts}
\usepackage{float}
\usepackage{hyperref}
\hypersetup{
  colorlinks = true,
  citecolor = blue,
  urlcolor = blue
}

\newcommand{{\bfl}}{\mbox{\boldmath$l$\unboldmath}}
\newcommand{{\bfx}}{\mbox{\boldmath$x$\unboldmath}}
\newcommand{{\bfk}}{\mbox{\boldmath$k$\unboldmath}}
\newcommand{{\bfv}}{\mbox{\boldmath$v$\unboldmath}}
\newcommand{{\bfq}}{\mbox{\boldmath$q$\unboldmath}}
\newcommand{{\bfp}}{\mbox{\boldmath$p$\unboldmath}}
\newcommand{{\bfa}}{\mbox{\boldmath$a$\unboldmath}}
\newcommand{{\bfT}}{\mbox{\boldmath$T$\unboldmath}}
\newcommand{{\bfE}}{\mbox{\boldmath$E$\unboldmath}}
\newcommand{{\bfm}}{\mbox{\boldmath$m$\unboldmath}}
\newcommand{{\bfe}}{\mbox{\boldmath$e$\unboldmath}}
\newcommand{{\bfc}}{\mbox{\boldmath$c$\unboldmath}}
\newcommand{{\bfg}}{\mbox{\boldmath$g$\unboldmath}}
\newcommand{{\bff}}{\mbox{\boldmath$f$\unboldmath}}
\newcommand{{\bfF}}{\mbox{\boldmath$F$\unboldmath}}
\newcommand{{\bfA}}{\mbox{\boldmath$A$\unboldmath}}
\newcommand{{\bfB}}{\mbox{\boldmath$B$\unboldmath}}
\newcommand{{\bfS}}{\mbox{\boldmath$S$\unboldmath}}
\newcommand{{\bfL}}{\mbox{\boldmath$L$\unboldmath}}
\newcommand{{\bfR}}{\mbox{\boldmath$R$\unboldmath}}
\newcommand{{\bfC}}{\mbox{\boldmath$C$\unboldmath}}
\newcommand{{\bfcalO}}{\mbox{\boldmath$\cal O$\unboldmath}}

\newcommand{{\bftheta}}{\mbox{\boldmath$\theta$\unboldmath}}
\newcommand{{\bfdelta}}{\mbox{\boldmath$\delta$\unboldmath}}
\newcommand{{\bfphi}}{\mbox{\boldmath$\phi$\unboldmath}}
\newcommand{{\bfrho}}{\mbox{\boldmath$\rho$\unboldmath}}
\newcommand{{\bfcalB}}{\mbox{\boldmath$\cal B$\unboldmath}}
\newcommand{{\bfcalE}}{\mbox{\boldmath$\cal E$\unboldmath}}
\newcommand{{\bfcalJ}}{\mbox{\boldmath$\cal J$\unboldmath}}

\newcommand{{\bfcalR}}{\mbox{\boldmath$\cal R$\unboldmath}}
\newcommand{{\bfbeta}}{\mbox{\boldmath$\beta$\unboldmath}}

\newcommand{\barlambda}{{\mkern0.75mu\mathchar '26\mkern -9.75mu\lambda}}

\def\v#1{{\bf#1}}

\begin{document}
\maketitle

\begin{abstract}
\noindent The Aharonov-Bohm (AB) effect was convincingly demonstrated using a micro-sized toroidal magnet but it is almost always explained using an infinitely-long solenoid or an infinitely-long flux line. The main reason for this is that the formal treatment of the AB effect considering a toroidal configuration turns out to be too cumbersome. But if the micro-sized toroidal magnet is modelled by a closed flux line of arbitrary shape and size then the formal treatment of the AB effect is exact, considerably simplified, and well-justified. Here we present such a treatment that covers in detail the electromagnetic, topological, and quantum-mechanical aspects of this effect. We demonstrate that the AB phase in a closed flux line is determined by a linking number and has the same form as the AB phase in an infinitely-long flux line which is determined by a winding number. We explicitly show that the two-slit interference shift associated with the AB effect in a closed flux line is the same as that associated with an infinitely-long flux line. We emphasise the topological nature of the AB phase in a closed flux line by demonstrating that this phase is invariant under deformations of the charge path, deformations of the closed flux line, simultaneous deformations of the charge path and the closed flux line, and the interchange between the charge path and the closed flux line. We also discuss the local and nonlocal interpretations of the AB effect in a closed flux line and introduce a non-singular gauge in which the vector potential vanishes in all space except on the surface surrounded by the closed flux line, implying that this vector potential is zero along the trajectory of the charged particle except on the crossing point where this trajectory intersects the surface bounded by the closed flux line, a result that questions the alleged physical significance of the vector potential and thereby the local interpretation of the AB effect.
\end{abstract}

\section{\large Introduction}
\label{1}
Quantum mechanics predicts that the wave function of a charged particle encircling an infinitely-long solenoid enclosing a uniform magnetic flux accumulates the AB phase \cite{1}. The charged particle moves in a non-simply connected region where there is no magnetic field and therefore there is no Lorentz force acting on the charge but there is a nonzero vector potential. The AB phase is topological because it depends on the winding number representing the number of times the charge carries out around the solenoid \cite{2}. This topological feature is manifested in the fact that this phase is independent of the dynamics of the encircling charge. The AB phase admits a nonlocal interpretation according to which the magnetic field of the solenoid acts on the charged particle in regions for which this field is excluded (see, for example, the textbook of Rohrlich and Aharonov \cite{2} for a representative view of this nonlocal interpretation). However, the most popular interpretation of this phase is that it is originated by the local action of the vector potential, in whose case this potential acquires a physical significance (see, for example, the textbook of Feynman \cite{3} for a representative view of this local interpretation). The AB phase is physically manifested in a modified two-slit interference experiment, in which a shift in the interference pattern proportional to the AB phase is observed. This is the AB effect. Regardless of its physical interpretation, the AB effect has become an influential effect in many branches of physics from condensed matter physics to high-energy particle physics, fluid mechanics, gravitation and cosmology (see, for example, Cohen et al. \cite{4}, and references therein).

On the experimental side, the first reports on the detection of the AB effect were due to Chambers \cite{5}, Fowler et al. \cite{6} and Boerch et al. \cite{7}, who used finite magnetic devices like magnetised whiskers and long solenoids. M\"ollenstedt and Bayh \cite{8} used a tiny solenoid with a diameter as small as one micron. The use of a finite solenoid raised questions by several authors (see, for example, Peshkin and Tonomura \cite{9}, and  Tonomura \cite{10}, and references therein) regarding the experimental verification of the AB effect by arguing that the electrons may not have been completely shielded from the magnetic field of the finite solenoid. But the concerns about the presence of a magnetic field in the electron paths
were removed by a series of experiments reported by Tonomura et al. \cite{11,12,13}. In a first instance, Tonomura et al. \cite{11} employed a squared micro-sized toroidal magnet whose leakage field effects were confined to be sufficiently small in order to verify the AB effect. In a second instance, Tonomura et al. \cite{12,13} used a circular micro-sized toroidal magnet covered with a superconducting layer which, due to the Meissner effect, essentially confined the magnetic field of the toroidal magnet. This allowed a more definite experimental verification of the AB effect.

But why being experimentally demonstrated  in a conclusive form using a toroidal configuration, is the AB effect generally explained using an idealised infinitely-long solenoid? We think that the answer deals with the fact that an exact treatment of the AB effect using a toroidal configuration and covering on equal footing its related electromagnetic, topological and quantum-mechanical features is a cumbersome task, which does not seem to have been reported so far. On the other hand, the posed question leads us to one of the more peculiar features of the AB effect: whenever the charged particle encircles a line of singularity, i.e. whenever it lies on a non-simply connected region, the existence of the AB effect does not depend on the particular geometry of the solenoid, which may be seen as a consequence of the topological character of this effect. Stated differently, in order for the wave function of the charged particle to accumulate the AB phase, the requirement of an idealised infinitely-long solenoid is sufficient but not necessary since this phase also arises by considering a less-idealised toroidal solenoid \cite{14,15,16}.

However, although in both an infinitely-long solenoid and a toroidal solenoid the AB effect arises, the mathematical treatment using the former solenoid is, as we have said before, considerably simpler than that using the latter solenoid because, among other reasons, the computation of the vector potential of the infinitely-long solenoid is much simpler compared to that of the toroidal solenoid \cite{17,18,19,20}. For this reason we think that the AB effect is generally explained considering an infinitely-long solenoid instead of a toroidal solenoid.

\begin{figure}
\centering
\includegraphics[scale=0.45]{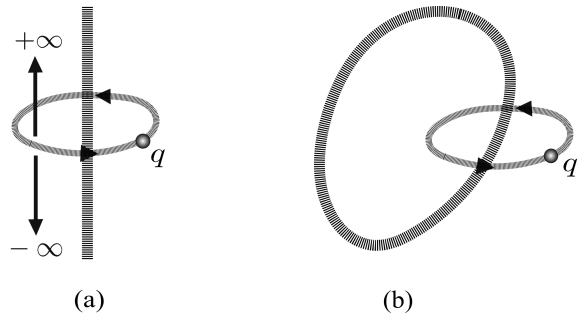}
\caption{\small{(a) A charge moving around an infinitely-long flux line. (b) A charge moving around a closed flux line. Both configurations are defined in non-simply connected regions in which the AB effect exists.}}
\label{Fig1}
\end{figure}

An infinitely-long flux line and a closed flux line of arbitrary shape are also electromagnetic configurations defined in non-simply connected regions (see Fig.~\ref{Fig1}). As expected, in both configurations the AB phase arises. However, while the AB phase in an idealised infinitely-long flux line has been extensively discussed, the AB phase in the less-idealised closed flux line has received much less attention and only a few authors have laterally addressed it \cite{21,22,23}. Since the more definite experimental verification of the AB effect relies on the use of a micro-sized superconducting toroidal magnet then the idea of modelling this magnet by a closed flux line of arbitrary shape and size is well justified.

In this paper, we present a detailed discussion of the AB effect in a closed flux line. The main purposes of this paper are the following: (i) to show that an exact and non-cumbersome treatment of the AB effect can be accomplished by considering a closed flux line of arbitrary shape, (ii) to emphasise the topological nature of the AB effect by showing that the AB phase arising in a closed flux line (determined by a linking number) has the same form than the AB phase arising in an infinitely-long flux line (determined by a winding number) and by demonstrating that the AB phase in a closed flux line is invariant under deformations of the charge path, deformations of the closed flux line, simultaneous deformations of the charge path and the closed flux line, and the interchange between the charge path and the closed flux line, (iii) to argue in favour of a nonlocal interpretation of the AB effect in a closed flux line by introducing a gauge that questions the local interpretation of this effect, and (iv) to discuss the difference in applying non-singular and singular gauge transformations in the
AB effect and show that the latter transformations modify the magnetic field.

Our paper is organised as follows. In Sec.~\ref{2} we discuss the formal aspects of the electrodynamics of a closed magnetised flux line. We derive the vector potential of a closed flux line, show that it may be expressed as the gradient of a multi-valued function, and prove that its circulation is gauge invariant. In Sec.~\ref{3} we show that the circulation of the vector potential is topological because it depends on a linking number and is nonlocal because it is delocalised with respect to the magnetic flux, a result arising from the Stokes theorem applied to the examined non-simply connected region. In Sec.~\ref{4} we derive the AB phase that accumulates the wave function of a charged particle upon continuously encircling the closed flux line and show that this phase exhibits the same form as the AB phase in an infinitely-long flux line modulo a linking number that specifies the former phase and a winding number that specifies the latter phase. In Sec.~\ref{5} we present a novel treatment of the AB two-slit interference experiment using a closed flux line and show that the corresponding shift detected on the second screen of the interference device coincides with that using an infinitely-long flux line. In Sec.~\ref{6} we introduce four topological invariances of the AB phase in a closed flux line that enlighten the topological nature of this phase. In Sec.~\ref{7} we discuss the local and nonlocal interpretations of the AB effect in a closed flux line. We argue in favour of the latter interpretation and against the former interpretation by stressing that the vector potential is gauge-dependent and its circulation is spatially delocalised. In Sec.~\ref{8} we strengthen our objection against the physical reality of the vector potential and the local interpretation of the AB effect by introducing a non-singular gauge in which the vector potential vanishes in all space except on the surface surrounded by the closed flux line which implies that as the charge encircles the closed flux line, the vector potential is zero along the trajectory of the charge except on a point of this trajectory. In Sec.~\ref{9} we discuss the subtle differences that exist in applying singular and non-singular gauge transformations in the AB effect and argue that only the latter transformations can consistently be applied in this effect because the former transformations modify the confined magnetic field. In Sec.~\ref{10} we summarise our main results. In Appendices A-E we demonstrate some equations relevant in our discussion.

\section{\large Vector potential of a closed flux line}
\label{2}
We can think of a closed flux line of arbitrary shape either as an infinitesimally thin closed magnetised solenoid or a closed line of magnetic dipoles. Any of these equivalent representations can be modelled using the steady electric current density
\begin{equation}
\v J = \frac{c\Phi}{4\pi} \nabla \times \oint_{\mathscr{C}}\delta( \v x- \v x')\, d \v x',
\end{equation}
where $\delta(\v x- \v x')$ is the Dirac delta function with $\v x$ being the field point and $\v x'$ the source point, $\Phi= 4\pi \lambda$ is the flux through the closed magnetised line with $\lambda$ being the magnetic dipole moment density per unit length, the line integral is evaluated along the closed flux line represented by the curve $\mathscr{C}$, the direction of the current is specified by the direction of the curve $\mathscr{C},$ and Gaussian units are adopted. The current satisfies $\nabla \cdot \v J = 0$ and is a magnetisation current $\v J = c\nabla \times \v M$ where
\begin{equation}
\v M= \frac{\Phi}{4\pi}\oint_{\mathscr{C}}\delta(\v x- \v x')\, d \v x',
\end{equation}
is the magnetisation vector confined along the curve $\mathscr{C}$. The associated magnetostatic equations are
\begin{equation}
\nabla \cdot \v B=0,\quad\nabla \times \v B=\Phi\nabla \times \oint_{\mathscr{C}}\delta(\v x- \v x')\, d \v x',
\end{equation}
whose solution is given by the magnetic field
\begin{equation}
\v B= \Phi\oint_{\mathscr{C}}\delta(\v x- \v x')\, d \v x',
\end{equation}
which is confined along the curve $\mathscr{C}$. From Eqs.~(2) and (4) it follows the relation $\v B = 4\pi \v M$ connecting the magnetic field with the magnetisation vector. To verify the homogeneous equation appearing in Eq.~(3), we use $\nabla \cdot \delta(\v x- \v x')d \v x'=\nabla\delta(\v x-\v x')\cdot d \v x'$ and $\nabla \delta(\v x- \v x')=-\nabla' \delta(\v x- \v x')$ so that $\nabla \cdot \v B= -\Phi\oint_{\mathscr{C}}\nabla'\delta(\v x- \v x')\cdot d\v x'=0$ which holds because $\oint_{\mathscr{C}}\nabla'\delta(\v x- \v x')\cdot d\v x'=0$ on account of the gradient theorem and the fact that $\delta(\v x- \v x')$ is a single-valued function of $\v x'$. The result $\oint_{\mathscr{C}}\nabla\delta(\v x- \v x')\cdot d \v x'=0$ has also been shown by Kleinert \cite{24,25}, deWit \cite{26}, and Kunin \cite{27}. From the homogeneous equation in Eq.~(3) it follows $\v B = \nabla \times \v A$ where $\v A$ is the associated vector potential. This relation, the inhomogeneous equation in Eq.~(3), the identity $\nabla^2 \v F = \nabla(\nabla \cdot\v F )-\nabla \times (\nabla \times \v F),$ and the adoption of the Coulomb gauge condition $\nabla \cdot \v A=0$, yield the Poisson equation
\begin{equation}
\nabla^2  \v A = - \Phi\nabla \times \oint_{\mathscr{C}}\delta(\v x- \v x')\, d \v x',
\end{equation}
whose solution is given by the vector potential
\begin{equation}
\v A=\frac{\Phi}{4\pi} \oint_{\mathscr{C}} \frac{(\v x' - \v x)\times d \v x'}{|\v x - \v x'|^3}.
\end{equation}
Using $\nabla \times (d \v x'/|\v x - \v x'|)= \nabla(1/|\v x- \v x'|)\times d \v x'$ and $\nabla(1/|\v x- \v x'|)=-\nabla'(1/|\v x- \v x'|)$ we can write Eq.~(6) in the following form
\begin{equation}
\v A=\frac{\Phi}{4\pi} \nabla\times \oint_{\mathscr{C}} \frac{d \v x'}{|\v x - \v x'|}.
\end{equation}
Considering $\nabla^2(\nabla\times \v F) =\nabla \times (\nabla^2\v F)$ and $\nabla^2(1/|\v x- \v x'|)=-4\pi \delta(\v x- \v x')$ we can verify that Eq.~(7) satisfies Eq.~(5): $\nabla^2\v A = (\Phi/4\pi)\nabla\times\oint_{\mathscr{C}}\nabla^2(d \v x'/|\v x - \v x'|)= -\Phi\nabla \times \oint_{\mathscr{C}}\delta(\v x- \v x')d \v x'$. Equation (7) satisfies the Coulomb gauge $\nabla\cdot \v A=0$ on account of $\nabla \cdot (\nabla \times \v F)=0$. In Appendix A we show that the curl of Eq.~(7) gives Eq.~(4) while in Appendix B  we show that Eq.~(7) can be written as
\begin{equation}
\v A = \frac{\Phi}{4\pi}\nabla\Omega_0 + \Phi \bm{\delta}_{\mathscr{S}},
\end{equation}
where $\Omega_0$ is the single-valued solid angle subtended by the curve $\mathscr{C}$ and defined by \cite{24,25}
\begin{equation}
\Omega_0 = \int_{\mathscr{S}} \frac{(\v x'- \v x) \cdot d \v S'}{|\v x- \v x'|^3},
\end{equation}
and $\bm{\delta}_{\mathscr{S}}$ is the Dirac delta surface vector function defined by \cite{24,25,26,27}
\begin{equation}
\bm{\delta}_{\mathscr{S}}= \int_{\mathscr{S}}\delta(\v x- \v x')\,d \v S',
\end{equation}
where $\mathscr{S}$ is the surface enclosed by the curve $\mathscr{C}$ and $d \v S'$ the differential surface vector normal to $\mathscr{S}$. Accordingly, the vector potential of the closed flux line can be expressed as the sum of a term involving the gradient of the single-valued solid angle $\nabla \Omega_0$ plus a term involving the delta function $\bm{\delta}_{\mathscr{S}}$, which is localised on the surface $\mathscr{S}$ enclosed by the curve $\mathscr{C}$. The function $\Omega_0$ is said to be single-valued because it satisfies in all space the Schwarz integrability condition, according to which the crossed second partial derivatives applied to the function $\Omega_0$ commute
\begin{equation}
(\partial^i \partial^j - \partial ^j \partial^i)\Omega_0=0,
\end{equation}
where index notation has been adopted in which summation on repeated indices is understood and $\partial^i=(\nabla)^i$ ---this condition  for the single valuedness of functions based on the Schwarz integrability condition is discussed in Kleinert's book on multi-valued fields \cite{24}. In Appendix C we prove the relations $\oint_C \nabla \Omega_0\cdot d \v x=0$ and $\nabla \times \nabla \Omega_0=0$, which allow us to demonstrate Eq.~(11). On the other hand, we can verify that the curl of the potential defined in Eq.~(8) gives the magnetic field specified in Eq.~(4). With this purpose we first consider the relation
\begin{equation}
\nabla \times \bm{\delta}_{\mathscr{S}}= \bm{\delta}_{\mathscr{C}},
\end{equation}
where $\bm{\delta}_{\mathscr{C}}$ is a vector line Dirac delta defined along the curve $\mathscr{C}$ and given by \cite{24,25,26,27}
\begin{equation}
\bm{\delta}_{\mathscr{C}}=\oint_{\mathscr{C}}\delta(\v x- \v x')\, d \v x'.
\end{equation}
Equation (12) has been mentioned by Kleinert \cite{24,25} and explicitly demonstrated by deWit \cite{26} and Kunin \cite{27}. A proof of Eq.~(12) goes as follows.  Using $\nabla \times [\delta(\v x- \v x')d \v S']=-d \v S'\times \nabla\delta(\v x- \v x')$ and $\nabla\delta(\v x- \v x') = -\nabla'\delta(\v x- \v x')$, the curl of Eq.~(10) gives $\nabla \times \bm{\delta}_{\mathscr{S}}=\int_{\mathscr{S}}d \v S'\times \nabla'\delta(\v x- \v x')$ and using the Stokes theorem $\int_{\mathscr{S}}d \v S'\times \nabla'\delta(\v x- \v x')= \oint_{\mathscr{C}}\delta(\v x- \v x')d \v x'$, where $\mathscr{C}$ is the boundary of $\mathscr{S}$, we obtain Eq.~(12). The curl of Eq.~(8) yields $\nabla \times \v A= [\Phi/(4\pi)] \nabla \times \nabla\Omega_0 + \Phi\nabla \times \bm{\delta}_{\mathscr{S}}$ and since $\Omega_0$ is single-valued then $\nabla \times \nabla \Omega_0=0$ so that $\nabla \times \v A= \Phi\nabla \times \bm{\delta}_{\mathscr{S}}$. Using Eq.~(12) we obtain  $\nabla \times \v A= \Phi\bm{\delta}_{\mathscr{C}}$, whose right-hand side identifies with the magnetic field in Eq.~(4).

Although the solid angle $\Omega_0(\v x)$ is a single-valued function, it is a discontinuous function as it jumps by $4\pi$ when the observation point $\v x$ crosses the surface $\mathscr{S}$ \cite{24,25}. This discontinuity has led several authors to the misconception that $\Omega_0$ is a multi-valued function (see, for example, Zangwill \cite{28}, Schwinger et al. \cite{29}, and Eyges \cite{30}). The fact that $\Omega_0$ is single-valued and discontinuous has been emphasised by Kleinert \cite{24,25} and has been explicitly demonstrated by Djuri\'c \cite{31}. We follow Kleinert \cite{24,25} and make use of the Schwarz integrability condition to prove in Appendix C that $\Omega_0$ is a single-valued function. We also note that the function $\Omega_0$ depends on the choice of the shape of the surface $\mathscr{S}$ \cite{24,25}. We can express Eq.~(8) in terms of the gradient of a multi-valued representation of the solid angle denoted as $\Omega$, which is continuous and independent of the choice of the surface $\mathscr{S}$. Kleinert \cite{24,25} has shown the following result:
\begin{equation}
\nabla \Omega = \nabla \Omega_0 + 4\pi\bm{\delta}_{\mathscr{S}}.
\end{equation}
\begin{figure}
	\centering
	\includegraphics[scale=0.55]{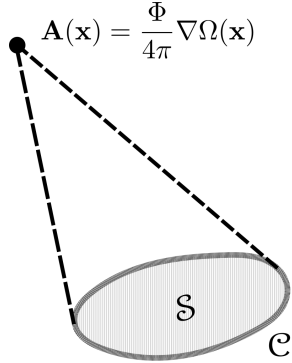}
	\caption{\small {Vector potential of a closed flux line evaluated at the point $\v x$. The solid angle is subtended by the surface $\mathscr{S}$ enclosed by the curve $\mathscr{C}$ representing the shape of the closed flux line.}}
	\label{Fig2}
\end{figure}
The function $\Omega$ is a multi-valued function because it violates the Schwarz integrability condition [i.e. $(\partial^i\partial^j-\partial^j \partial^i)\Omega \neq 0].$ Kleinert has pointed out the result \cite{24,25}
\begin{equation}
(\partial^i \partial^j - \partial^j \partial^i)\Omega=4\pi \varepsilon^{ijk}(\bm{\delta}_{\mathscr{C}})_k,
\end{equation}
where $\varepsilon^{ijk}$ is the Levi-Civita symbol and $(\bm{\delta}_{\mathscr{C}})_k$ denotes the components of Eq.~(13). In Appendix D we explicitly demonstrate Eq.~(15). Using Eq.~(15) we can express Eq.~(8) in the useful form
\begin{equation}
\v A= \frac{\Phi}{4\pi}\nabla\Omega.
\end{equation}
A pictorial description of the vector potential given by Eq.~(16) is shown in Fig.~\ref{Fig2}. Since Eq.~(16) can be written as the gradient of a function then we would have $\nabla \times \v A=0$ and hence the vanishing of the magnetic field. Indeed, this is the case in all space except along $\mathscr{C}$. In fact, multiplying Eq.~(15) by $\varepsilon_{mij}$ we have $\varepsilon_{mij}(\partial^i \partial^j - \partial^j \partial^i)\Omega=4\pi\varepsilon_{mij} \varepsilon^{ijk}(\bm{\delta}_{\mathscr{C}})_k,$ which implies $2\varepsilon_{mij}\partial^i \partial^j\Omega=8\pi(\bm{\delta}_{\mathscr{C}})_m$ and therefore $(\nabla\times\nabla\Omega)_m
=4\pi(\bm{\delta}_{\mathscr{C}})_m$, or equivalently
\begin{equation}
\nabla\times\nabla\Omega=4\pi\bm{\delta}_{\mathscr{C}},
\end{equation}
which gives $\nabla\times\v A= [\Phi/(4\pi)]\nabla\times\nabla\Omega=\Phi\bm{\delta}_{\mathscr{C}}$ in agreement with Eq.~(4). We should note that multi-valued functions, such as $\Omega$ in Eq.~(16), are characteristic in the electrodynamics of flux lines or magnetised strings which are defined in non-simply connected regions \cite{24,25,32,33}. Two further examples of multi-valued functions appear in the following configurations: the vector potential outside an infinitely-long flux line in cylindrical coordinates \cite{1}: $\v A= \Phi \,\hat{\!\bfphi}/(2\pi \rho)$ and the vector potential of the Dirac monopole in spherical coordinates \cite{34}: $\v A_{D} =g (1- \cos\theta)\,\hat{\!\bfphi}/(r\sin\theta),$ where $g$ is the magnetic charge. The former vector potential may be expressed as $\v A=\nabla\chi$ where $\chi=\Phi\phi/(2\pi)$ is a multi-valued function that satisfies $(\partial^{i}\partial^{j}-\partial^{j}\partial^{i})\chi \neq0$. The latter vector potential satisfies $\nabla \cdot(\nabla \times \v A_{D})\neq0$ implying \cite{24}: $(\partial^{i}\partial^{j}-\partial^{j}\partial^{i})(\v A_{D})_k \neq 0,$ which makes the Dirac monopole potential $\v A_{D}$ a multi-valued function. Discussions on electromagnetic aspects of the Dirac monopole can be found in Kleinert \cite{24,25}, Heras \cite{35}, and Shnir \cite{36}.

Let us now discuss the gauge invariance of the circulation of the vector potential of the closed flux line. The circulation $\oint_C \v A\cdot d \v x$ taken along an arbitrary closed path $C$ is invariant under the gauge transformation $\v A'=\v A+\nabla\Lambda$ where $\Lambda$ is the corresponding gauge function. The gauge transformation must be a non-singular gauge transformation, i.e. one in which the gauge function $\Lambda$
is a single-valued function satisfying the Schwarz integrability condition \cite{24,25} $(\partial^i\partial^j-\partial^j \partial^i)\Lambda=0$. Moreover, this condition should hold in all space and not only within a finite region. Therefore,
\begin{equation}
\oint_C \v A' \cdot d \v x= \oint_C \v A \cdot d \v x+ \oint_C \nabla\Lambda \cdot d \v x= \oint_C \v A \cdot d \v x,
\end{equation}
which follows because $\oint_{C}\nabla \Lambda \cdot d \v x=0$ on account of the single-valuedness of $\Lambda.$ Let us insist that Eq.~(18) holds whenever
$\Lambda$ is a single-valued function. If $\Lambda$ were a multi-valued function then we would have $\oint_{C}\nabla\Lambda \cdot d \v x\neq 0$ and this would imply $\oint_{C}\v A'\cdot d \v x\neq \oint_{C}\v A\cdot d \v x$ and therefore the breaking of the gauge invariance of the circulation of the vector potential.

We also note that if the Coulomb gauge condition in the transformed potential $\nabla \cdot \v A'=0$ is preserved then the corresponding gauge function, in addition to be a single-valued function, it should be a restricted gauge function satisfying $\nabla^2\Lambda=0$. Nevertheless, this additional requirement is not necessary for the validity of Eq.~(18) since $\v A'$ need not be in the Coulomb gauge as we will see in Sec.~\ref{8} of this paper.

\section{\large Topology and nonlocality of the circulation of the vector potential}
\label{3}
\begin{figure}
	\centering
	\includegraphics[scale=0.63]{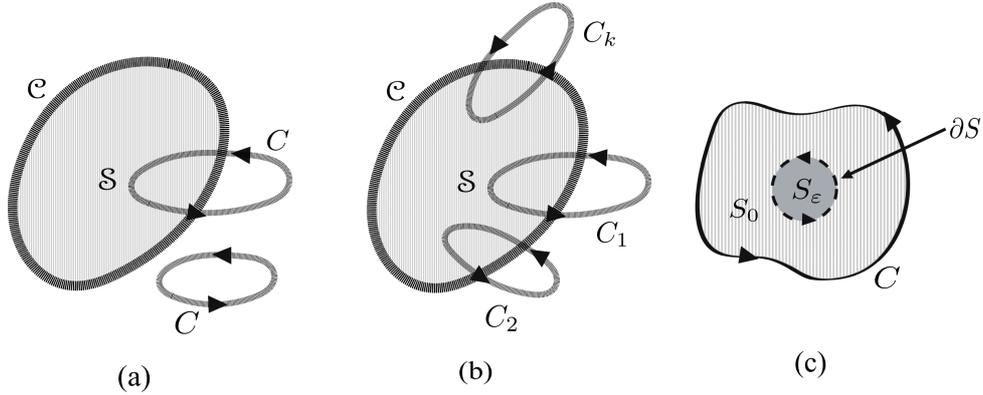}
	\caption{\small{(a) If the circulation of the vector potential $\v A$ along the path $C$ encircles the closed flux line of shape $\mathscr{C}$ then this configuration is defined in a non-simply connected region and $\oint_{C} \v A \cdot d \v x =l\Phi$ where $l$ is the linking number of the paths $C$ and $\mathscr{C}$. If the circulation of the vector potential $\v A$ along the path $C$ does not encircle the closed flux line of shape $\mathscr{C}$ then the configuration is simply-connected and $\oint_{C} \v A \cdot d \v x =0.$ (b) The circulation of the vector potential $\v A$ along the different closed paths $C_1, C_2, ...,C_k$ encircling the closed flux line of shape $\mathscr{C}$ and corresponding to the same linking number $l$ are equivalent: $\oint_{C_1}\v A\cdot d\v x = \oint_{C_2}\v A\cdot d\v x=...=\oint_{C_k}\v A\cdot d \v x$. (c) The circulation $\oint_{C}\v A \cdot d \v x$ is taken along a path $C$ greater than the boundary $\partial S_{\varepsilon}$ of the infinitesimal surface $S_{\varepsilon}$ pierced by the closed flux line. Since $\oint_{C>\partial S_{\varepsilon}}\v A \cdot d \v x= \int_{S_{\varepsilon}} \v B\cdot d \v S$ holds then the circulation of the vector potential is spatially delocalised from the surface where the magnetic flux is non-vanishing.}}
	\label{Fig3}
\end{figure}

The space containing a closed flux line is non-simply connected because there is a non-removable line of singularity along the curve $\mathscr{C}$ where the magnetic field is confined. This non-simply connected space implies some interesting topological and nonlocal features of the circulation of the vector potential of the closed flux line which will now be discussed. Let us first discuss the topological aspect. The circulation of the vector potential given by Eq.~(6) along an arbitrary closed path $C$ gives
\begin{equation}
\oint_{C}\v A\cdot d \v x = \frac{\Phi}{4\pi} \oint_{C}  \oint_{\mathscr{C}} \frac{[(\v x' - \v x)\times d \v x']\cdot d \v x}{|\v x - \v x'|^3},
\end{equation}
where it is assumed that the path $C$ does not intersect the curve $\mathscr{C}$ where the closed flux line is defined. Using $[(\v x' - \v x)\times d \v x']\cdot d \v x=(\v x - \v x')\cdot (d\v x \times d\v x')$ in Eq.~(19) we obtain
\begin{equation}
\oint_{C}\v A\cdot d \v x = \Phi \bigg[ \frac{1}{4\pi} \oint_{C}  \oint_{\mathscr{C}} \frac{(\v x - \v x')\cdot (d\v x \times d\v x')}{|\v x - \v x'|^3}\bigg].
\end{equation}
The quantity within the brackets is identified in the general case with the Gauss linking number $l$ (or linking integral) which is defined as \cite{17,24,37}
\begin{equation}
\frac{1}{4\pi} \oint_{C}  \oint_{\mathscr{C}} \dfrac{(\v x - \v x')\cdot (d\v x \times d\v x')}{|\v x - \v x'|^3}=
\begin{cases}
l(C,\mathscr{C})& \text{if $C$ encloses $\mathscr{C}$} \\ 0 &\text{otherwise}
\end{cases}
\end{equation}
and represents the number of times the path $C$ encloses the curve $\mathscr{C}.$ The linking number can be positive or negative depending on the direction of $C$ and $\mathscr{C}$ and whether if $C$ crosses above or below the surface $\mathscr{S}$ bounded by $\mathscr{C}$ as projected in a two-dimensional plane. From this result it follows that changing the direction of $C$ and $\mathscr{C}$ will change the sign of the linking number, i.e. $l(-C,\mathscr{C})= -l(C,\mathscr{C})$ and $l(C,-\mathscr{C})= -l(C,\mathscr{C}).$ Moreover changing simultaneously the direction of $C$ and $\mathscr{C}$ leaves the linking number invariant: $l(-C,-\mathscr{C})= l(C,\mathscr{C}).$ Inserting Eq.~(21) in Eq.~(20) it follows
\begin{equation}
\oint_{C}\v A\cdot d \v x =
\begin{cases}
l\Phi & \text{if $C$ encloses $\mathscr{C}$} \\ 0 &\text{otherwise}
\end{cases}
\end{equation}
which shows the topological nature of the circulation of the vector potential of the closed flux line: if the path $C$ encloses the closed flux line of shape $\mathscr{C}$ then this configuration is non-simply connected and the circulation of the vector potential $\v A$ accumulates $l$ times the magnetic flux $\Phi$. The product $l\Phi$ is a constant quantity and therefore it is independent of the path $C$. If the path $C$ does not encircle the closed flux line of shape $\mathscr{C}$ then the configuration is locally simply connected and the circulation of $\v A$ vanishes [see Fig.~\ref{Fig3} (a)]. By locally simply connected we mean here any finite region of space that does not include the closed flux line, i.e. the closed line of singularity $\mathscr{C}$. In such regions $\bm{\delta}_{\mathscr{S}}=\int_{\mathscr{S}} \delta(\v x- \v x')d\v S'=0$ because $\v x$ is not on any point on the surface $\mathscr{S}$ and therefore Eq.~(8) takes the form $\v A(\v x \notin \mathscr{S}) = [\Phi/(4\pi)]\nabla\Omega_0$ whose circulation vanishes because $\Omega_0$ is a single-valued function.
A comment on the interpretation of the path $C$ and the curve $\mathscr{C}$ in Eq.~(20) is pertinent. The path $C$ identifies with the closed trajectory on which is defined the circulation of the vector potential $\v A$ while the curve $\mathscr{C}$ identifies with the closed magnetic flux line which is assumed to be stationary. The physical quantity that can be identified along the direction of the curve $\mathscr{C}$ is the steady current $\v J$ defined in Eq.~(1) which flows continuously within the closed flux line. In short: $C$ is the path of the circulation of $\v A$ while $\mathscr{C}$ is the closed flux line whose direction is followed by its steady current $\v J$.

Equation~(14) can be used to obtain two convenient ways to define the linking number, which in turn can be applied to the circulation of the vector potential of the closed flux line. We first observe that Eq.~(16) implies $\nabla \Omega = 4\pi\v A/\Phi$ which combines with Eq.~(6) to give the relation
\begin{equation}
\nabla\Omega=\oint_{\mathscr{C}} \frac{(\v x' - \v x)\times d \v x'}{|\v x - \v x'|^3},
\end{equation}
whose circulation reads
\begin{equation}
\oint_C\nabla\Omega\cdot d \v x=\oint_{C}  \oint_{\mathscr{C}} \frac{(\v x - \v x')\cdot (d\v x \times d\v x')}{|\v x - \v x'|^3},
\end{equation}
where we have used $[(\v x' - \v x)\times d \v x']\cdot d \v x=(\v x - \v x')\cdot (d\v x \times d\v x')$. The right-hand side of Eq.~(24) is equal to $4\pi$ times the linking number specified in Eq.~(21) and therefore it follows that
\begin{equation}
\frac{1}{4 \pi}\oint_{C}  \nabla\Omega\cdot d\v x=
\begin{cases}
l& \text{if $C$ encloses $\mathscr{C}$} \\ 0 &\text{otherwise}
\end{cases}
\end{equation}
which shows that the circulation of the gradient of the multi-valued solid angle along a path $C$ enclosing $\mathscr{C}$ is non-vanishing and proportional to the linking number. On the other hand, Eq.~(14) tells us that $\nabla \Omega = \nabla \Omega_0 + 4\pi\bm{\delta}_{\mathscr{S}}$ which is used in the left-hand side of Eq.~(25) to obtain
\begin{equation}
\frac{1}{4 \pi}\oint_{C}  \nabla\Omega_0\cdot d\v x +\oint_{C}\bm{\delta}_{\mathscr{S}}\cdot d \v x =
\begin{cases}
l& \text{if $C$ encloses $\mathscr{C}$} \\ 0 &\text{otherwise}
\end{cases}
\end{equation}
But $\oint_{C}\nabla\Omega_0 \cdot d\v x=0$ because $\Omega_0$ is a single-valued function and therefore
\begin{equation}
\oint_{C}  \bm{\delta}_{\mathscr{S}}\cdot d \v x=
\begin{cases}
l& \text{if $C$ crosses $\mathscr{S}$} \\ 0 &\text{otherwise}
\end{cases}
\end{equation}
which shows that the circulation of the Dirac surface vector $\bm{\delta}_{\mathscr{S}}$ is non-vanishing when the closed path $C$ crosses the surface $\mathscr{S}$. Kleinert \cite{24,25} has discussed Eqs.~(25) and (27). The use of Eq.~(25) in Eq.~(22) gives
\begin{equation}
\oint_{C}\v A\cdot d \v x = \frac{\Phi}{4\pi}\oint_{C}\nabla\Omega \cdot d \v x=
\begin{cases}
l\Phi & \text{if $C$ encloses $\mathscr{C}$ } \\ 0 &\text{otherwise}
\end{cases}
\end{equation}
while the use of Eq.~(27) in Eq.~(22) gives
\begin{equation}
\oint_{C}\v A\cdot d \v x =
\Phi\oint_{C}\bm{\delta}_{\mathscr{S}}\cdot d \v x=
\begin{cases}
l\Phi & \text{if $C$ crosses $\mathscr{S}$ } \\ 0 &\text{otherwise}
\end{cases}
\end{equation}
Equation (28) tells us that the circulation of $\v A$ is proportional to the circulation of $\nabla\Omega$ which is non-vanishing when the path $C$ encircles the curve $\mathscr{C}$. Equation (29) tells us that the circulation of $\v A$ is non-vanishing and proportional to the circulation of $\bm{\delta}_{\mathscr{S}}$ along any path $C$ crossing the surface $\mathscr{S}$ ---which enlightens the fact that the path $C$ is irrelevant as long as it crosses the surface $\mathscr{S}$. Since the circulation of $\v A$ is insensitive to the path $C$ then we can consider $C_1,C_2...C_k$ different paths each one encircling the closed flux line and corresponding to the same linking number. Therefore,
\begin{equation}
\oint_{C_1}\v A \cdot d \v x = \oint_{C_2}\v A \cdot d \v x=...=\oint_{C_k}\v A \cdot d \v x.
\end{equation}
The paths $C_1,C_2...C_k$ are homotopically equivalent and therefore we could not distinguish if the flux $\Phi$ is connected with the circulation of $\v A$ along $C_1$ or along $C_2$ or along  $C_k$ [see Fig.~\ref{Fig3} (b)]. This indistinguishability is a manifestation of the topology of the circulation of the vector potential of the closed flux line which lies in a non-simply connected region.

Let us now discuss the nonlocal aspect of the circulation of the vector potential. We can apply the Stokes theorem to the left-hand side of Eq.~(22) but we need to be careful in doing so when the path $C$ encloses the closed flux line of shape $\mathscr{C}$
because this circulation is defined in a non-simply connected region. Applying the Stokes theorem we obtain
\begin{equation}
\oint_{C=\partial S} \v A\cdot d \v x = \int_{S} \nabla \times \v A\cdot d \v S,
\end{equation}
where $\partial S$ is the boundary of the total surface $S$ enclosed by the path $C.$ We can write this total surface as $S= S_0 + S_{\varepsilon}$, where $S_0$ is the surface that excludes the closed flux line and $S_{\varepsilon}$ is the infinitesimal surface pierced by the closed flux line [see Fig.~\ref{Fig3} (c)]. Of course, $C= \partial S >\partial S_{\varepsilon}$ because the path $C$ encircles the closed flux line. The notation $C>\partial S_{\varepsilon}$ states that the length of the curve $C$ is greater than the length of the boundary $\partial S_{\varepsilon}$ of the infinitesimal surface $S_{\varepsilon}$ pierced by the closed flux line. It then follows that Eq.~(31) takes the form
\begin{equation}
\oint_{C=\partial S} \v A\cdot d \v x = \int_{S_0} \nabla \times \v A\cdot d\v S + \int_{S_{\varepsilon}}\nabla \times \v A \cdot d \v S.
\end{equation}
Since the surface $S_0$ excludes the closed flux line then the first term on the right-hand side vanishes $\int_{S_0} \nabla\times \v A\cdot d\v S=0$ because $\nabla \times \v A =0$ along the surface $S_0$. Therefore, the Stokes theorem gives
\begin{equation}
\oint_{C>\partial S_{\varepsilon}} \v A\cdot d \v x = \int_{S_{\varepsilon}} \nabla \times \v A \cdot d \v S,
\end{equation}
which expresses a nonlocal relation: while the circulation $\oint_{C>\partial S_{\varepsilon}} \v A\cdot d \v x$ is defined outside the closed flux line, the flux $\int_{S_{\varepsilon}} \nabla \times \v A \cdot d \v S$ is defined on the infinitesimal surface $S_{\varepsilon}$ pierced by the closed line of singularity $\mathscr{C}$, i.e. the sides of Eq.~(33) are defined in different regions of space implying a nonlocal connection between them. This nonlocality is a consequence of having applied the Stokes theorem in a non-simply connected region. To have a better physical picture of Eq.~(33) we can write in the right-hand side the magnetic field $\nabla \times \v A = \v B$ which is non-vanishing along the infinitesimal surface $S_{\varepsilon}$. This gives the relation
\begin{equation}
\oint_{C>\partial S_{\varepsilon}} \v A\cdot d \v x = \int_{S_{\varepsilon}} \v B \cdot d \v S,
\end{equation}
and this shows that the circulation of the vector potential evaluated outside the closed flux line is delocalised with respect to the flux of the magnetic field confined along $\mathscr{C}$. In other words, there is a nonlocal relation between the circulation of $\v A$ and the flux of $\v B$.

By applying the Stokes theorem to the vector $\bm{\delta}_{\mathscr{S}}$, it follows that
\begin{equation}
\oint_{C=\partial S} \bm{\delta}_{\mathscr{S}}\cdot d \v x=\int_{S}\nabla \times \bm{\delta}_{\mathscr{S}}\cdot d \v S,
\end{equation}
where $C$ is the boundary of the surface $S$. Using Eq.~(35) and the relation $\nabla \times \bm{\delta}_{\mathscr{S}}= \bm{\delta}_{\mathscr{C}}$ given in Eq.~(12), we obtain
\begin{equation}
\oint_{C=\partial S} \bm{\delta}_{\mathscr{S}}\cdot d \v x=\int_{S} \bm{\delta}_{\mathscr{C}}\cdot d \v S,
\end{equation}
which connects the circulation of the vector $\bm{\delta}_{\mathscr{S}}$ with the surface integral of the vector $\bm{\delta}_{\mathscr{C}}$. Equations (36) and Eq.~(29) give an alternative definition of the linking number
\begin{equation}
\int_{S} \bm{\delta}_{\mathscr{C}} \cdot d\v S=
\begin{cases}
l& \text{if $\mathscr{C}$ crosses S} \\ 0 &\text{otherwise}
\end{cases}
\end{equation}
The surface in Eq.~(37) can be written as $S=S_0 + S_{\varepsilon}$ where $S_0$ is the surface not pierced by the closed line of singularity  $\mathscr{C}$ while $S_{\varepsilon}$ is the infinitesimal surface pierced by the curve $\mathscr{C}$. Thus
\begin{equation}
\int_{S_0} \bm{\delta}_{\mathscr{C}} \cdot d\v S + \int_{S_{\varepsilon}} \bm{\delta}_{\mathscr{C}} \cdot d\v S=
\begin{cases}
l& \text{if $\mathscr{C}$ crosses S} \\ 0 &\text{otherwise}
\end{cases}
\end{equation}
But  $\int_{S_0} \bm{\delta}_{\mathscr{C}} \cdot d\v S=0$ because  $\bm{\delta}_{\mathscr{C}}=0$ along $S_0$ ($\mathscr{C}$ never crosses $S_0$) and therefore Eq.~(38) reduces to
\begin{equation}
\int_{S_{\varepsilon}} \bm{\delta}_{\mathscr{C}} \cdot d\v S=
\begin{cases}
l& \text{if $\mathscr{C}$ crosses $S_{\varepsilon}$} \\ 0 &\text{otherwise}
\end{cases}
\end{equation}
where $S_{\varepsilon}\in S$. Equations (37) and (39) are equivalent representations of the linking number with the former being discussed by Kleinert \cite{25}. Using Eq.~(39) and the relation $\Phi\bm{\delta}_{\mathscr{C}}=\v B$ it follows that
\begin{equation}
\int_{S_{\varepsilon}} \v B \cdot d \v S=
\begin{cases}
l\Phi& \text{if $\mathscr{C}$ crosses $S_{\varepsilon}$} \\ 0 &\text{otherwise}
\end{cases}
\end{equation}
and this shows that the quantity $\int_{S_{\varepsilon}} \v B \cdot d \v S$ is the accumulated magnetic flux along the closed line of singularity $\mathscr{C}$. Both Eq.~(22) and Eq.~(40) are equivalent and both are in agreement with Eq.~(34). Now, from Eqs.~(30) and (34) we obtain the relation
\begin{equation}
\oint_{C_k>\partial S} \v A\cdot d \v x=...=\oint_{C_2>\partial S_{\varepsilon}} \v A\cdot d \v x=\oint_{C_1>\partial S_{\varepsilon}} \v A\cdot d \v x=\int_{S_{\varepsilon}} \v B \cdot d \v S.
\end{equation}
in which the equalities on the left-hand side express a manifest ambiguity because we cannot distinguish if the flux $\int_{S_{\varepsilon}} \v B\cdot d \v S$ is connected with the circulation of $\v A$ along $C_1>\partial S$ or along $C_2>\partial S$, or along  $C_k>\partial S_{\varepsilon}$. In other words, the circulations in Eq.~(41) are delocalised with respect to the magnetic flux, an expected result since they are not functions of point. We should emphasise the difference between applying the Stokes theorem in a simply connected region and in the non-simply connected region considered here. In the former application of the theorem there is only a single curve $C=\partial S$ representing the boundary $\partial S$ of the surface $S$. In the latter application of the theorem there can be $k$ curves $C_k>\partial S$ all of them greater (i.e. delocalised) than the infinitesimal boundary $\partial S_{\varepsilon}$ of the surface $S_{\varepsilon}$ pierced by the closed line of singularity $\mathscr{C}$. Interestingly, similar features can be found in other electromagnetic configurations defined in non-simply connected regions. In two recent papers \cite{38,39} we have shown that a similar relation to that in Eq.~(41) holds in the circulations of vector potentials define in three non-simply connected configurations: (i) the circulation of the magnetic vector potential outside an infinitely-long magnetic solenoid, (ii) the circulation of the electric vector potential outside an infinitely-long electric solenoid, and (iii) the circulation of the sum of the magnetic and electric vector potentials outside an infinitely-long dual solenoid which confines its electric and magnetic fields. The fact that the circulations of vector potential of different geometrical configurations are delocalised enlightens their topological character.

We can now draw the lessons we have learned so far about the peculiarities of the electromagnetism of a closed flux line. The magnetisation current $\v J$ in Eq.~(1) yields the vector potential $\v A$ in Eq.~(6) and the magnetic field $\v B$ in Eq.~(4). The closed flux line involves a closed line of singularity $\mathscr{C}$ and then one can apply the definition of the Gauss linking number [Eq.~(21)] and the Stokes theorem [Eq.~(31)] in this non-simply connected region. Both mathematical tools lead to Eq.~(41) which unambiguously shows the nonlocality of the circulation of $\v A$ with respect to the flux of the confined magnetic field $\v B.$ One can then argue that the background of this nonlocality is of topological nature. To see this we use Eqs.~(27) and (39) to obtain the beautiful topological relation
\begin{equation}
\oint_{C>\partial S_{\varepsilon}} \bm{\delta}_{\mathscr{S}}  \cdot d\v x=l=\int_{S_{\varepsilon}} \bm{\delta}_{\mathscr{C}} \cdot d\v S,
\end{equation}
which contains all the geometrical information about the nonlocality of the electromagnetic relation in Eq.~(34). Let us emphasise that Eq.~(42) is a purely topological quantity devoid of any physical content. When multiplied by the flux $\Phi$, Eq.~(42) yields the electromagnetic relation in Eq.~(34) ---see, Eqs.~(22) and (40). After seeing these results, we cannot avoid saying that \emph{topology dictates nonlocality!}

\section{\large AB phase in a closed flux line}
\label{4}
\begin{figure}
	\centering
	\includegraphics[scale=0.63]{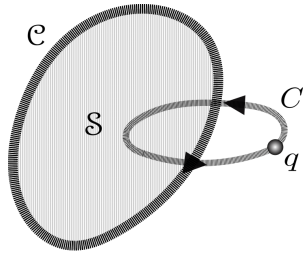}
	\caption{\small{As the charged particle encircles the closed flux line along the path $C$ its wave function accumulates the AB phase $\delta=ql\Phi/(\hbar c)$ where $l$ in the linking number representing the number of times $C$ encircles the closed flux line of shape $\mathscr{C}$ and which encloses the surface $\mathscr{S}$.}}
	\label{Fig4}
\end{figure}

Consider a non-relativistic particle of mass $m$ and charge $q$ that is continuously moving around the closed flux line (see Fig.~\ref{Fig4}). The corresponding time-dependent Schr\"odinger equation is given by
\begin{equation}
i \hbar \frac{\partial \Psi}{\partial t}=\frac{1}{2m}\bigg(\!-i\hbar\nabla- \frac{q}{c}\v A \bigg)^2\Psi,
\end{equation}
where $\v A$ is the vector potential of the closed flux line. Since $\v A$ can be written as the gradient of a function as seen in Eq.~(16) then a solution of Eq.~(43) can be obtained by multiplying the free solution $\Psi_0$ ---which satisfies Eq.~(43) when $\v A=0$--- by a suitable local phase factor
\begin{equation}
\Psi(\v x,t) = {\rm e}^{[i q/(\hbar c)]\int_{\Gamma}\v A(\v x')\cdot d \v x'}\,\Psi_0(\v x,t),
\end{equation}
where the line integral in the phase is taken along the charge path $\Gamma$ from a fixed reference point $\small\bfcalO$ to the variable point $\v x$. We assume that the points $\small\bfcalO$ and $\v x$ never lie on the closed flux line while the path $\Gamma$ never crosses it. As the charge continuously encircles the closed flux line, it follows that any charge path $\Gamma$ can be decomposed as $\Gamma=C+\gamma$ where $C$ is any closed path that accounts for the number of times the charge encircles the closed flux line and $\gamma$ is any non-closed path that accounts for the open trajectory that the charge takes before completing another turn around the closed flux line. Thus, we can write
\begin{equation}
\int_{\Gamma}\v A\cdot d \v x'=\oint_{C}\v A\cdot d \v x' + \int_{\gamma}\v A\cdot d \v x'.
\end{equation}
Using this relation the wave function in Eq.~(44) takes the form $\Psi= {\rm e}^{[i q/(\hbar c)][\oint_{C}\v A\cdot d \v x' + }$ ${}^{ \int_{\gamma}\v A\cdot d \v x']}\Psi_0.$ As the path $C$ encircles the closed flux line of shape $\mathscr{C}$ it follows from Eq.~(22) that $\oint_{C}\v A\cdot d \v x'=l\Phi$ and thus,
\begin{equation}
\int_{\Gamma}\v A\cdot d \v x'=l\Phi + \int_{\gamma}\v A\cdot d \v x'.
\end{equation}
Using Eqs.~(44) and (46) we can write
\begin{equation}
\Psi(\v x,t) = [{\rm e}^{ilq\Phi/(\hbar c)}]\,{\rm e}^{[i q/(\hbar c)]\int_{\gamma}\v A(\v x')\cdot d \v x'}\,\Psi_0(\v x,t),
\end{equation}
which states that after the charged particle takes $l$ turns around the closed flux line, its wave function picks up the phase factor ${\rm e}^{ilq\Phi/(\hbar c)}$ and thereby it accumulates the AB phase
\begin{equation}
\delta = l\frac{q\Phi}{\hbar c},
\end{equation}
where $l$ is the linking number of the charge path $C$ around the closed flux line of shape $\mathscr{C}$. A first observation is that the derived AB phase in a closed flux line has the same form than the AB phase in an infinitely-long flux line localised along the $z$-axis. In fact, the latter phase is given by $\delta=n q\Phi/(\hbar c)$ where $n$ is the winding number representing the number of times the charge encircles the infinitely-long flux line. Thus, the only difference between the AB phase in a closed flux line and the AB phase in an infinitely-long flux line is the linking number that specifies the former phase and the winding number that specifies the latter phase. However, both AB phases are equivalent in the sense that the topological numbers $l$ and $n$ have the same interpretation in these phases: the number of times a charged particle encircles a line of singularity (closed or infinitely-long). The fact that the AB phase arises in different geometrical configurations is a manifestation of its topological nature.

We can write the AB phase in Eq.~(48) in other different forms each of which enlightens some of its properties. The form in Eq.~(48) shows that the AB phase in a closed flux line is topological because it depends on the linking number $l$ and is independent of the dynamics of the encircling charge, i.e. independent of the charge path $C$. A manifestly gauge-invariant form of this phase is obtained by using Eqs.~(22) and (48),
\begin{equation}
\delta =\frac{q}{\hbar c}\oint_C \v A \cdot d\v x,
\end{equation}
which is gauge invariant on account of the gauge invariance of the circulation of the vector potential [see Eq.~(18)]. Another form of the AB phase is obtained by using Eqs.~(40) and (48):
 \begin{equation}
 \delta = \frac{q}{\hbar c}\int_{S_{\varepsilon}} \v B \cdot d \v S.
 \end{equation}
 This form of the AB phase is conceptually interesting because it clearly admits a nonlocal interpretation: while the charge $q$ is moving along the path $C$ defined outside the closed flux line, the flux $\int_{S_{\varepsilon}} \v B \cdot d \v S$ is defined along the infinitesimal surface $S_{\varepsilon}$ where the magnetic field $\v B$ is non-vanishing. This suggests a nonlocal interaction between the electric charge and the magnetic field confined along the closed flux line. Another form of the AB phase may be obtained by inserting the flux $\Phi = 4\pi \lambda$ in Eq.~(48):  $\delta= 4\pi l q \lambda/(\hbar c),$ according to which the AB phase may be seen as the result of the nonlocal interaction between the electric charge $q$ moving outside the closed flux line and the magnetic dipole moment linear density $\lambda$ confined along the closed flux line. Another interesting form of the AB phase can be obtained using Eq.~(25) and Eq.~(48):
\begin{equation}
\delta = \frac{q\Phi}{4 \pi \hbar c}\oint_{C}\nabla\Omega \cdot d \v x,
\end{equation}
where $\Omega$ is the multi-valued solid angle defined by Eq.~(14). This form shows that the AB phase in a closed flux line has a high range of validity since Eq.~(51) is coordinate-independent. This fact shows that the AB phase is a quite general phenomenon that occurs whenever a charged particle encircles a magnetised closed flux line. From Eqs.~(27) and (48) we can obtain another form of the AB phase,
\begin{equation}
\delta = \frac{q\Phi}{\hbar c}\oint_{C}\bm{\delta}_{\mathscr{S}}\cdot d \v x,
\end{equation}
which is interesting in the sense that it explicitly shows that the AB phase in a closed flux line is not continuously accumulated as
the charged particle travels the path $C$ but discretely picked up as the charged particle crosses the surface $\mathscr{S}$ along which the function $\bm{\delta}_{\mathscr{S}}$ is non-vanishing. This result is independent of the shape of the surface $\mathscr{S}$.

\section{\large Quantum interference}
\label{5}
The AB phase is physically manifested in a modified two-slit interference experiment for charged particles in which an infinitely-long solenoid that confines its magnetic flux is placed between the two screens in the interference device. The presence of the solenoid causes a shift in the corresponding interference pattern which is proportional to the AB phase. This experiment has been discussed in early treatments of the AB effect (see, for example, Feynman's textbook \cite{3}). However, a detailed theoretical treatment of this interference effect has not been done for the case of the closed flux line.

Consider a double-slit interference experiment in which identical charged particles propagate from a source, pass into the two slits of a first screen, and are finally detected on the second screen. The computation of the corresponding wave function is somewhat laborious but we can take advantage of a similar  calculation given by Kobe \cite{40}. Therefore, we will follow arguments similar to those given by Kobe \cite{40}. The following assumptions are made: (i) Charged particles are localised in the $x$-$y$ plane and their motion perpendicular to the screens is treated classically while their motion parallel to the screens is treated quantum-mechanically. This assumption is valid if the velocity of the charged particles parallel to the screens is sufficiently high \cite{40}. Under this condition, the charges will follow classical paths along the $y$-axis. (ii) Charged particles are emitted with constant velocity $v$ in the $y$-direction which is perpendicular to the screens and have random velocity along the $x$-axis. This will allow us to use a one-dimensional wave function along the $x$-axis to describe the charged particles. (iii) The slits are infinitely-long in the $z$-direction and are Gaussian slits instead of rectangular slits. This will not change our main conclusions but will help in simplifying calculations \cite{40}.

In order to obtain the wave function for points on the second screen let us first consider the case in which there is only one slit localised at $+x_0$ and the closed flux line is absent. A charged particle propagates at the time $0$ from the origin to a point $x_a$ at the later time $t_a$ on the first screen localised a distance $y_a=v t_a$ where $v$ is the constant velocity along the $y$-axis. The charge then passes through a slit centred at $x_a=+x_0$ and continues to propagate until reaching a variable point $x_b$ on the second screen at the later variable time $t_b$ located a distance $y_b=vt_b$. The time taken for the charge to travel from $x_a$ to $x_b$ is then $t_b-t_a$. The corresponding wave function at $x_b$ and $t_b$ should then satisfy the one-dimensional time-dependent Schr\"odinger equation
\begin{equation}
i\hbar \frac{\partial\Psi_0}{\partial t_b}= -\frac{\hbar^2}{2m}\frac{\partial^2\Psi_0}{\partial x^{2}_{b}}.
\end{equation}
Kobe \cite{40} has calculated the corresponding wave function
\begin{eqnarray}
\nonumber \Psi_{0}(x_b, t_b)= \sqrt{\frac{m}{2 \pi i \hbar[t_b  + i\alpha t_a(t_b -t_a)]}}\qquad\qquad\qquad\qquad\qquad\qquad\,\,\,\,\\
\qquad\qquad\qquad\qquad\times \exp\bigg[ -\frac{(1-i\chi)(x_b - v_0t_b)^2}{2(\Delta x)^2} +i\beta \frac{(x_b-x_0)^2}{(t_b-t_a)} + i\beta\frac{x^{2}_{0}}{t_a}\bigg],
\end{eqnarray}
where $\beta=m/2\hbar$, $\alpha = \hbar/mb^2$ in which $b$ corresponds to a rectangular slit of width $\sqrt{\pi}b$, $v_0 = x_0 /t_a$ is the average velocity of the charge in the $x-$direction, $(\Delta x)^2=(b t_b/t_a)^2 + [\hbar(t_b-t_a)/mb]^2$ is the square of the total broadening where $(b t_b/t_a)$ is the classical Gaussian broadening arising due to the propagation of the particle from the origin to the slit at time $t_a$ and from this slit to the observation point on the second screen at the later time $t_b-t_a$ and $\hbar(t_b-t_a)/mb$ is the quantum-mechanical broadening due to the uncertainty of the momentum of the charged particle along the $x$-axis $m\Delta v_x = \hbar/b$ in going through a slit of width $b$, and $\chi=(b t_b/t_a)/[\hbar(t_b-t_a)/mb]$ is the ratio of the classical to the quantum-mechanical broadening. Let us now carry out the two-slit interference experiment in which the first slit is localised at $+x_0$ and the second slit is localised at $-x_0$. From the principle of superposition the wave function can be written as $\Psi_0 = \Psi^{1}_{0} + \Psi^{2}_{0},$ where $\Psi^{0}_{1}$ is the partial wave passing through the first slit centred at $+x_a$ and $\Psi^{0}_{2}$ is the partial wave passing through the second slit centred at $-x_a$. The partial wave $\Psi^{1}_{0}$ is of the same form as Eq.~(54) while the partial wave $\Psi^{2}_{0}$ can be obtained by making the replacements $+x_0\to-x_0$ and $v_0 \to - v_0$ in Eq.~(54). Therefore
\begin{eqnarray}
\nonumber \Psi_{0}(x_b, t_b)=\sqrt{\frac{m}{2 \pi i \hbar[t_b  + i\alpha t_a(t_b -t_a)]}}\exp\bigg[i\beta\frac{x^{2}_{0}}{t_a}\bigg]\qquad\qquad\qquad\qquad\\
\nonumber \times\bigg\{ \exp\bigg[ -\frac{(1-i\chi)(x_b - v_0t_b)^2}{2(\Delta x)^2} +i\beta \frac{(x_b-x_0)^2}{(t_b-t_a)}\bigg] \\
\qquad\qquad\qquad\qquad+ \exp\bigg[ -\frac{(1-i\chi)(x_b + v_0t_b)^2}{2(\Delta x)^2} +i\beta \frac{(x_b+x_0)^2}{(t_b-t_a)}\bigg] \bigg\}.
\end{eqnarray}
Using Eq.~(55) and doing some simplifications, we obtain the probability density
\begin{eqnarray}
\nonumber |\Psi_{0}(x_b,t_b)|^2=\frac{m}{4\pi^2\hbar^2[t^{2}_{b} \!+\! \alpha^2 t^{2}_{a}(t_b\!-\!t_a)^2]}\bigg\{\! \exp\bigg[\!-\! \frac{(x_b\!-\!v_0 t_b)^2}{(\Delta x)^2} \bigg] \!+\! \exp\bigg[\!-\! \frac{(x_b\!+\!v_0 t_b)^2}{(\Delta x)^2} \bigg] \\
\nonumber + 2\exp\bigg[-\frac{[(x_b-v_0t_b)^2 +(x_b+v_0t_b)^2]}{2(\Delta x)^2}\bigg]\qquad\qquad\qquad\qquad\qquad\qquad\\
\qquad\times \cos\bigg[ \frac{\chi [(x_b\!-\! v_0t_b)^2 -(x_b\!+\!v_0t_b)^2]}{2(\Delta x)^2} + \frac{\beta[(x_b \!-\! x_0)^2-(x_b\!+\!x_0)^2]}{(t_b-t_a)}\bigg]\bigg\}.\quad
\end{eqnarray}
In order for the interference effect to be significant it is necessary that \cite{40}: $v_0 t_b>>\Delta x,$ i.e. that the total broadening $\Delta x$ should be much larger than the average distance traveled by the charge in the $x$-direction through one of the slits. Under this condition
we can write $(x_b-v_0 t_b)^2/(\Delta x)^2\approx x^{2}_{b}/(\Delta x)^2$ and $(x_b+v_0 t_b)^2/(\Delta x)^2\approx x^{2}_{0}/(\Delta x)^2$ and thus the probability density reduces to
\begin{equation}
|\Psi_{0}(x_b,t_b)|^2=\frac{m\,{\rm e}^{-x^{2}_{0}/(\Delta x)^2}}{\pi^2\hbar^2[t^{2}_{b} + \alpha^2 t^{2}_{a}(t_b-t_a)^2]}\cos(\delta_0/2),
\end{equation}
where $\delta_0=[(2m x_b x_0) / \hbar (t_a-t_b)]$ is the phase angle. Using the de Broglie relation $\barlambda=\hbar/p$ where $p=m v_y$ is the constant momentum along the $y$-axis and letting $d=2x_0$ be the distance between the two slits and $L = v(t_a-t_b)=y_a-y_b$ the distance in the $y$-direction between the two screens, it follows that $\delta_0 =d x_b/L\barlambda$ and therefore a point $x_b$ detected on the second screen reads
\begin{equation}
x_b= \frac{L \barlambda}{d}\delta_0.
\end{equation}
 Let us now repeat the two-slit interference experiment but this time with the closed flux line
\begin{figure}
	\centering
	\includegraphics[scale=0.29]{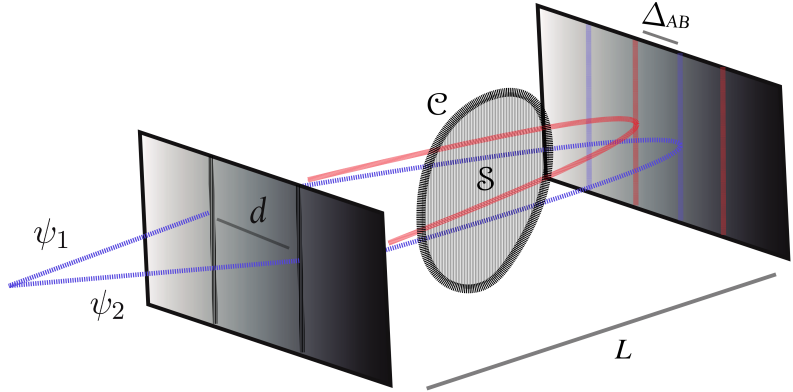}
	\caption{\small{Two-slit interference setup. Blue lines indicate partial wave packets that pass through the slits when the closed flux line is absent while red lines indicate partial wave packets that pass through the slits when the closed flux line is inserted between the two screens. The presence of the closed flux line gives rise to the additional shift $\Delta_{AB}$ detected on the second screen.}}
	\label{Fig5}
\end{figure}
of shape $\mathscr{C}$ inserted within the two slits as seen in Fig.~\ref{Fig5}. As in the case without the closed flux line, we first consider the solution when there is one slit only localised at the point $x_a=+x_0$. The magnetic field vanishes outside the closed flux line but the vector potential is non-vanishing. Therefore we are interested in solving the time-dependent Schr\"odinger equation
\begin{equation}
i\hbar \frac{\partial\Psi}{\partial t_b}= \frac{1}{2m}\bigg(-i\hbar \frac{\partial}{\partial x_b}- \frac{q}{c}A_{x}\bigg)^2\Psi,
\end{equation}
where $A_{x}=A_{x}(x_b,y_b,0)$ is the $x$ component of the vector potential evaluated at $x_b$, $y_b,$ and $0$. Since the vector potential is the gradient of a scalar function $\v A = [\Phi/(4\pi)]\nabla \Omega$ as seen in Eq.~(16) then we can follow the same approach as that in Eq.~(43) but applied to our one-dimensional problem. We multiply $\Psi_0$ in Eq.~(54) times a suitable phase factor and obtain
\begin{equation}
\Psi(x_b,t_b)={\rm e}^{[iq/(\hbar c)] \int^{(x_b,y_b,0)}_{0}\v A(\v x')\cdot d \v x'}\Psi_{0}(x_b,t_b),
\end{equation}
where the path of the line integral goes from the origin, passes through the slit centred at $x_a=+x_0$, and ends at $x_b,y_b,0,$ with $x_b$ being a variable point. Considering the two slits it follows that we can write $\Psi = \Psi_{1} + \Psi_{2}$ where $\Psi_1$ is the partial wave going through the first slit and $\Psi_2$ is the partial wave going through the second slit. We can then write $\Psi_1= {\rm e}^{[iq/(\hbar c)]\int_{1}\v A\cdot d \v x'}\Psi^{1}_{0}$ and $\Psi_2= {\rm e}^{[iq/(\hbar c)]\int_{2}\v A\cdot d \v x'}\Psi^{2}_{0}$ where $\int_1$ denotes the path of the line integral that goes from the origin, passes through the first slit centred at $+x_0$, and ends at $x_b,y_b,0$, while $\int_2$ denotes the path of the line integral that goes from the origin, passes through the second slit centred at $-x_0$, and ends at $x_b,y_b,0$. Therefore we can write
\begin{equation}
\Psi(x_b,t_b) = [\Psi^{1}_{0}(x_b,t_b) + {\rm e}^{[iq/(\hbar c)][\int_2 \v A(\v x') \cdot d \v x' -\int_1 \v A(\v x') \cdot d \v x' ]}\Psi^{2}_{0}(x_b,t_b)]{\rm e}^{[iq/(\hbar c) ]\int_1 \v A(\v x)\cdot d \v x'}.
\end{equation}
Assuming that the difference of the trajectories $1$ and $2$ forms a closed path $C=2-1$ encircling the closed flux line once then it follows from Eq.~(22) that the phase difference in Eq.~(61) becomes
\begin{equation}
\frac{q}{\hbar c}\bigg[\int_2\v A(\v x')\cdot d \v x'-\int_1\v A(\v x')\cdot d \v x'\bigg]= \frac{q}{\hbar c}\oint_{C=2-1}\v A(\v x')\cdot d \v x'=\frac{q\Phi}{\hbar c},
\end{equation}
where the closed path $C=2-1$ corresponds to the linking number $l=1$ around the closed flux line of shape $\mathscr{C}$ and we assume that the direction of $\mathscr{C}$ is properly oriented so that the linking number is positive. Using Eq.~(62) the wave function in Eq.~(61) becomes
\begin{equation}
\Psi(x_b,t_b) = [\Psi^{1}_{0}(x_b,t_b) + {\rm e}^{iq\Phi/(\hbar c)}\Psi^{2}_{0}(x_b,t_b)]{\rm e}^{[iq/(\hbar c) ]\int_1 \v A(\v x')\cdot d \v x'}.
\end{equation}
The partial wave $\Psi^{1}_{0}$ is of the same form as Eq.~(54) while the partial wave $\Psi^{2}_{0}$ can be calculated by making the replacements $+x_0\to-x_0$ and $v_0 \to - v_0$ in Eq.~(54). After some simplifications, the probability density becomes
\begin{eqnarray}
\nonumber |\Psi(x_b,t_b)|^2=\frac{m}{4\pi^2\hbar^2[t^{2}_{b} \!+\! \alpha^2 t^{2}_{a}(t_b\!-\!t_a)^2]}\bigg\{ \exp\bigg[\!-\! \frac{(x_b\!-\!v_0 t_b)^2}{(\Delta x)^2} \bigg] \!+\! \exp\bigg[\!-\! \frac{(x_b\!+\!v_0 t_b)^2}{(\Delta x)^2} \bigg]\\
\nonumber + 2\exp\bigg[-\frac{[(x_b-v_0t_b)^2 +(x_b+v_0t_b)^2]}{2(\Delta x)^2}\bigg] \qquad\qquad\qquad\qquad\quad\qquad\quad\qquad\qquad\\
\times \cos\bigg[ \frac{\chi [(x_b- v_0t_b)^2 -(x_b+v_0t_b)^2]}{2(\Delta x)^2} + \frac{\beta[(x_b - x_0)^2-(x_b+x_0)^2]}{(t_b-t_a)} - \frac{q\Phi}{\hbar c} \bigg]\bigg\}.\quad
\end{eqnarray}
Since the AB phase $q\Phi/(\hbar c)=\delta$ is generally not a multiple of $2\pi$ (this case will be shortly addressed) it follows that there is a non-vanishing interference effect attributed to the this phase which should manifest in the two-slit interference effect. The total interference will be non-negligible provided $v_0 t_b>>\Delta x$ and under this condition we can approximate $(x_b-v_0 t_b)^2/(\Delta x)^2\approx x^{2}_{b}/(\Delta x)^2$ and $(x_b+v_0 t_b)^2/(\Delta x)^2\approx x^{2}_{0}/(\Delta x)^2$, and therefore the probability density in Eq.~(64) reduces to
\begin{equation}
|\Psi(x_b,t_b)|^2=\frac{m\,{\rm e}^{-x^{2}_{0}/(\Delta x)^2}}{\pi^2 \hbar^2[t^{2}_{b} + \alpha^2 t^{2}_{a}(t_b-t_a)^2]}\cos(\delta'/2),
\end{equation}
where $\delta'=[(2m x_b x_0) / \hbar (t_a-t_b)] + (q\Phi/\hbar c)$ is the phase angle. Using $\barlambda=\hbar/p$ and letting $d=2x_0$ be the distance between the two slits and $L = v_y(t_a-t_b)=y_b-y_a$ the distance in the $y$-direction between the two screens, it follows that $\delta' =d [x_b + L\barlambda q \Phi/(d\hbar c)]/(L\barlambda)$. Therefore the effect of the closed flux line physically manifests in a shift of the observation point $x_b \to x_b+ \Delta_{AB},$ where
\begin{equation}
\Delta_{AB}= \frac{L\barlambda}{d}\frac{q \Phi}{\hbar c},
\end{equation}
denotes the AB shift. A first observation is that the derived AB shift in a closed flux line is the same as the AB shift corresponding to an infinitely-long flux line \cite{40}. This result enlightens the topological nature of the AB effect since $\Delta_{AB}$ arises in different geometrical configurations. On the other hand, Eq.~(66) can be interpreted as a nonlocal effect: the shift $\Delta_{AB}$ explicitly depends on the flux $\Phi$ but it is evaluated in a point on the second screen where this flux is zero. This suggests that the shift $\Delta_{AB}$ manifests a nonlocality in which the particles of charge $q$ are affected by the magnetic field in a region for which this field vanishes.

We can write Eq.~(66) in other different forms each one which enlightens some of its properties. Inserting the flux $\Phi=4\pi \lambda$ in Eq.~(66) it follows that
\begin{equation}
\Delta_{AB}= 4\pi\frac{L\barlambda}{d}\frac{q \lambda}{\hbar c},
\end{equation}
which admits a nonlocal interpretation according to which the shift $\Delta_{AB}$ originates from the nonlocal interaction between the electric charge $q$ and the magnetic dipole moment linear density $\lambda.$ Equation (66) can be written in terms of the AB phase $\delta= q\Phi/(\hbar c)$, i.e.
\begin{equation}
\Delta_{AB}= \frac {L \barlambda}{d} \delta.
\end{equation}
Using Eqs.~(51) and (68) together with the de Broglie relation \,$\barlambda=\hbar/p$ we obtain
\begin{equation}
\Delta_{AB} = \frac{Lq\Phi}{4\pi pcd}\oint_{C=2-1}\nabla \Omega\cdot d \v x.
\end{equation}
This equation shows that the AB shift $\Delta_{AB}$ depends on the circulation of the gradient of the multi-valued representation of the solid angle $\Omega$ defined in Eq.~(14). This result is independent of the path $C$, i.e. independent of the dynamics of the charged particles having the momentum $p$ ---interestingly, Eq.~(69) may be interpreted as a classical relation in the sense that it explicitly involves only classical pieces. Another form of $\Delta_{AB}$ follows from using  Eqs.~(52) and (68) together with \,$\barlambda=\hbar/p$,
\begin{equation}
\Delta_{AB} = \frac{Lq\Phi}{4\pi pcd}\oint_{C=2-1}\bm{\delta}_{\mathscr{S}}\cdot d \v x.
\end{equation}
As may be seen, it only suffices for one of the partial wave packets to cross the surface $\mathscr{S}$ along the path $1$ while the other to surround it along the path $2$ to produce the interference shift in Eq.~(70) (see Fig.~\ref{Fig5}). This result is independent of the paths of the partial waves or the surface $\mathscr{S}.$

Let us now discuss the case in which the interference effect reflected in the shift in Eq.~(66) is non-observable. This happens when we assume that the flux through the closed flux line is quantised, i.e. when $\Phi= N \Phi_0$ where $N$ is an integer and $\Phi_0=2\pi \hbar c/e$ is the flux quantum with $e$ being the electron's charge. Moreover the electric charge is quantised $q=n_e e$ with $n_e$ being an integer and  therefore the probability density reads $|\Psi|^2=|\Psi^{1}_{0} + {\rm e}^{i q \Phi/(\hbar c)}\Psi^{2}_{0}|^2=|\Psi^{1}_{0} + \Psi^{2}_{0}|^2$ where ${\rm e}^{i q \Phi/(\hbar c)}={\rm e}^{i2\pi n_eN}=1$ on account of $n_eN$ being an integer. It then follows that any interference effect due to the magnetic flux becomes unobservable. This is reflected in the corresponding shift $x_b+\Delta_{AB}=x_b+n_eN2\pi L \barlambda/d$ which is equivalent to Eq.~(58) when the closed flux line is absent. In other words, shifting the phase angle $\delta_0\to \delta_0 + 2\pi n_e N$ does not produce any observable effect because the spectral lines detected on the second screen of the two-slit device merely get relabelled but are otherwise unchanged. In the experiments of Tonomura \cite{12,13} et al. a micro-sized ferromagnetic toroidal magnet covered with a superconducting layer was used. Characteristically, in superconducting rings the flux is quantised in the form $\Phi=N\Phi_0/2$ where the factor $2$ arises due to the corresponding Cooper pairs in the superconductor. If in addition we only consider electrons $q=e$ then it follows that $|\Psi|^2=|\Psi^{1}_{0} + {\rm e}^{i q \Phi/(\hbar c)}\Psi^{2}_{0}|^2=|\Psi^{1}_{0} + {\rm e}^{i \pi N}\Psi^{2}_{0}|^2.$ Thus, if $N$ is even then ${\rm e}^{i \pi N}=1$ and there is no AB effect but if $N$ is odd then ${\rm e}^{i \pi N}=-1$ and there is the AB effect. In the practice, Tonomura et al. \cite{12,13} did not use the two-slit setup discussed here but a novel interference method based on electron holography \cite{10,11,12,13}. However, the success of Tonomura et al. \cite{12,13} relied on verifying that when $N$ was odd then the AB effect was detected.

\section{\large Topological invariances of the AB phase in a closed flux line}
\label{6}
In this section, we will introduce four topological invariances of the AB phase in a closed flux line. Let $C$ be the charge path and $\mathscr{C}$ the shape of the closed flux line. The following topological invariances of the AB phase in a closed flux line follow:
\vskip 4pt
\noindent  \textbf{(a) Deformations of the charge path.} The AB phase in Eq.~(48) is invariant under deformations of the charge path from the initial path $C$ to the final path $C'$. This invariance is expressed by
\begin{equation}
\delta(C,\mathscr{C})=\delta(C',\mathscr{C});\quad \{ C\longrightarrow C'\},
\end{equation}
where $\delta(C,\mathscr{C})$ is the AB phase corresponding to the charge path $C$ around the closed flux line of shape $\mathscr{C}$ and $\delta(C',\mathscr{C})$ is the AB phase corresponding to the deformed charge path $C'$ around the closed flux line. Both phases can be written as
\begin{eqnarray}
\nonumber \delta(C,\mathscr{C}) = \frac{q\Phi}{4\pi \hbar c}\oint_{C}  \oint_{\mathscr{C}} \dfrac{(\v x - \v x')\cdot (d\v x \times d\v x')}{|\v x - \v x'|^3},\\
\delta(C',\mathscr{C}) = \frac{q\Phi}{4\pi \hbar c}\oint_{C'}  \oint_{\mathscr{C}} \dfrac{(\v x - \v x')\cdot (d\v x \times d\v x')}{|\v x - \v x'|^3}.
\end{eqnarray}
To demonstrate Eq.~(71) we apply the transformation $C\to C'$ to the quantity
\begin{equation}
\frac{1}{4\pi}\oint_{C}  \oint_{\mathscr{C}} \dfrac{(\v x - \v x')\cdot (d\v x \times d\v x')}{|\v x - \v x'|^3},
\end{equation}
and obtain
\begin{equation}
\frac{1}{4\pi}\oint_{C}  \oint_{\mathscr{C}} \dfrac{(\v x - \v x')\cdot (d\v x \times d\v x')}{|\v x - \v x'|^3}=\frac{1}{4\pi}\oint_{C'}  \oint_{\mathscr{C}} \dfrac{(\v x - \v x')\cdot (d\v x \times d\v x')}{|\v x - \v x'|^3},
\end{equation}
which is explicitly proved in Appendix E. We also note that the relation in Eq.~(74) has been demonstrated by Gelca \cite{41}. The use of Eq.~(74) implies Eq.~(71) because of Eq.~(72).
\vskip 4pt
\noindent \textbf{(b) Deformations of the closed flux line.} The AB phase in Eq.~(48) is invariant under deformations of the closed flux line from the initial shape $\mathscr{C}$ to the final shape $\mathscr{C}'$
\begin{equation}
\delta(C,\mathscr{C})=\delta(C,\mathscr{C}');\quad \{\mathscr{C}\longrightarrow \mathscr{C}'\},
\end{equation}
where $\delta(C,\mathscr{C})$ is the AB phase corresponding to the charge path $C$ around the closed flux line of shape $\mathscr{C}$ and $\delta(C,\mathscr{C}')$ is the AB phase corresponding to the charge path around the closed flux line of deformed shape $\mathscr{C}'$. These phases can be written as
\begin{eqnarray}
\nonumber \delta(C,\mathscr{C}) = \frac{q\Phi}{4\pi \hbar c}\oint_{C}  \oint_{\mathscr{C}} \dfrac{(\v x - \v x')\cdot (d\v x \times d\v x')}{|\v x - \v x'|^3},\\
\delta(C,\mathscr{C}') = \frac{q\Phi}{4\pi \hbar c}\oint_{C}  \oint_{\mathscr{C}'} \dfrac{(\v x - \v x')\cdot (d\v x \times d\v x')}{|\v x - \v x'|^3}.
\end{eqnarray}
To demonstrate Eq.~(75) we apply the transformation $\mathscr{C}\to \mathscr{C}'$ to Eq.~(73) and obtain
\begin{equation}
\frac{1}{4\pi}\oint_{C}  \oint_{\mathscr{C}} \dfrac{(\v x - \v x')\cdot (d\v x \times d\v x')}{|\v x - \v x'|^3}=\frac{1}{4\pi}\oint_{C}  \oint_{\mathscr{C}'} \dfrac{(\v x - \v x')\cdot (d\v x \times d\v x')}{|\v x - \v x'|^3},
\end{equation}
a result proved Appendix E. The use of Eq.~(77) implies Eq.~(75) because of Eq.~(76).
\vskip 4pt
\noindent \textbf{(c) Deformations of the charge path and the closed flux line.} The AB phase in Eq.~(48) is invariant under simultaneous deformations of the charge path from $C$ to $C'$ and the shape of the closed flux line from $\mathscr{C}$ to $\mathscr{C}'.$ This invariance is expressed by the transformation
\begin{equation}
\delta(C,\mathscr{C})=\delta(C',\mathscr{C}');\quad \{ C \longrightarrow C',\mathscr{C} \longrightarrow\mathscr{C}'\},
\end{equation}
where $\delta(C,\mathscr{C})$ is the AB phase corresponding to the charge path $C$ around the closed flux line of shape $\mathscr{C}$ and $\delta(C',\mathscr{C}')$ is the AB phase corresponding to the deformed charge path $C'$ around the deformed closed flux line of shape $\mathscr{C}'$ These two AB phases can be written in the following form
\begin{eqnarray}
\nonumber \delta(C,\mathscr{C}) = \frac{q\Phi}{4\pi \hbar c}\oint_{C}  \oint_{\mathscr{C}} \dfrac{(\v x - \v x')\cdot (d\v x \times d\v x')}{|\v x - \v x'|^3},\,\\\delta(C',\mathscr{C}') = \frac{q\Phi}{4\pi \hbar c}\oint_{C'}  \oint_{\mathscr{C}'} \dfrac{(\v x - \v x')\cdot (d\v x \times d\v x')}{|\v x - \v x'|^3}.
\end{eqnarray}
To demonstrate Eq.~(78) we apply the transformations $C\to C'$ and $\mathscr{C}\to \mathscr{C}'$ to Eq.~(73) and obtain
\begin{equation}
\frac{1}{4\pi}\oint_{C}  \oint_{\mathscr{C}} \dfrac{(\v x - \v x')\cdot (d\v x \times d\v x')}{|\v x - \v x'|^3}=\frac{1}{4\pi}\oint_{C'}  \oint_{\mathscr{C}'} \dfrac{(\v x - \v x')\cdot (d\v x \times d\v x')}{|\v x - \v x'|^3},
\end{equation}
which is proved in Appendix E. The use of Eq.~(80) implies Eq.~(78) because of Eq.~(79).
\vskip 4pt
\noindent \textbf{(d) Interchange between the charge path and the closed flux line}. The AB phase in Eq.~(48) is invariant under the simultaneous interchange between the charge path $C$ and the closed flux line of shape $\mathscr{C}.$ This interchange is represented by the transformation
\begin{equation}
\delta(C, \mathscr{C})= \delta(\mathscr{C},C); \quad \{C \longleftrightarrow\mathscr{C}\},
\end{equation}
where the AB phases $\delta(C,\mathscr{C})$ and $\delta(\mathscr{C}, C)$ can be written in the following form
\begin{eqnarray}
\nonumber \delta(C,\mathscr{C}) = \frac{q\Phi}{4\pi \hbar c}\oint_{C}  \oint_{\mathscr{C}} \dfrac{(\v x - \v x')\cdot (d\v x \times d\v x')}{|\v x - \v x'|^3},\\
\delta(\mathscr{C},C) = \frac{q\Phi}{4\pi \hbar c}\oint_{\mathscr{C}}\oint_{C}   \dfrac{(\v x - \v x')\cdot (d\v x \times d\v x')}{|\v x - \v x'|^3}.
\end{eqnarray}
The proof of Eq.~(81) follows from the well-known relation \cite{37,41}
\begin{equation}
\frac{1}{4\pi}\oint_{C}  \oint_{\mathscr{C}} \dfrac{(\v x - \v x')\cdot (d\v x \times d\v x')}{|\v x - \v x'|^3}= \frac{1}{4\pi}\oint_{\mathscr{C}}\oint_{C}  \dfrac{(\v x - \v x')\cdot (d\v x \times d\v x')}{|\v x - \v x'|^3},
\end{equation}
which implies Eq.~(81) because of Eq.~(82).

\begin{figure}[h]
	\centering
	\includegraphics[scale=0.65]{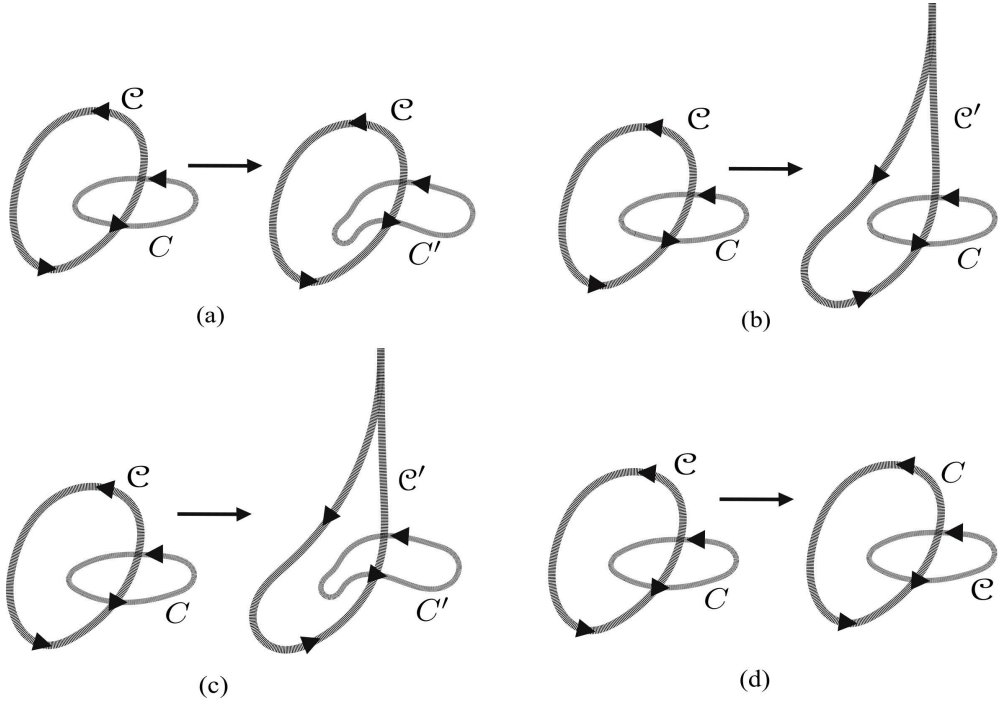}
	\caption{\small{(a) The AB phase is invariant under deformations of the charge path. (b) The AB phase is invariant under deformations of the closed flux line. (c) The AB phase is invariant under simultaneous deformations of the charge path and the closed flux line. (d) The AB phase is invariant under the interchange between the charge path and the closed flux line.}}
	\label{Fig6}
\end{figure}

As may be seen, the AB phase in a closed flux line is a truly topological quantity which is manifested in the fact that this phase is invariant under the topological transformations given in Eqs. (71), (75), (78), and (81) (a pictorial description of these topological invariances is shown in Fig.~\ref{Fig6}). The reason for these topological invariances is simple: the linking number $l(C,\mathscr{C})$ of the curves $C$ and $\mathscr{C}$ defined in Eq.~(21) is invariant under deformations of $C,$ deformations of $\mathscr{C},$ simultaneous deformations of $C$ and $\mathscr{C},$ and the simultaneous interchange between $C$ and $\mathscr{C}.$ Accordingly,
\begin{equation}
l(C,\mathscr{C})=l(C',\mathscr{C}),\,\,\, l(C,\mathscr{C})=l(C,\mathscr{C}'), \,\,\, l(C,\mathscr{C})=l(C',\mathscr{C}'),\,\,\, l(C,\mathscr{C})=l(\mathscr{C},C),
\end{equation}
which follow from Eqs.~(74), (77), (80), and (83). Since the AB phase in a closed flux line is proportional to the linking number then the topological invariances of the linking number are translated into the AB phase. Moreover, any quantity proportional to the AB phase will also share these topological invariances. For example, consider the AB shift in Eq.~(66) which can be written as $\Delta_{AB}=L \barlambda \delta(C,\mathscr{C})/d$ and which corresponds to $l=1.$ Considering Eq.~(84) it follows that
\begin{equation}
\Delta_{AB}= \frac{L\barlambda}{d}\delta(C,\mathscr{C})= \frac{L\barlambda}{d}\delta(C',\mathscr{C})=\frac{L\barlambda}{d}\delta(C,\mathscr{C}')=\frac{L\barlambda}{d}\delta(C',\mathscr{C}')=\frac{L\barlambda}{d}\delta(\mathscr{C},C).
\end{equation}
The second equality in Eq.~(85) shows that the shift $\Delta_{AB}$ is invariant under deformations of the path $C=2-1$ formed by the difference of the trajectories of the partial wave packets $\Psi_1$ and $\Psi_2$ in the two-slit interference effect. The third equality in Eq.~(85) shows that the shift $\Delta_{AB}$ is insensitive to deformations of the closed flux line of shape $\mathscr{C}.$ The fourth equality in Eq.~(85) shows that the shift $\Delta_{AB}$ is invariant under simultaneous deformations of the path $C$ and the closed flux line of shape $\mathscr{C}.$ The fifth equality in Eq.~(85) shows that the shift $\Delta_{AB}$ is invariant under the simultaneous interchange of the path $C$ and the closed flux line of shape $\mathscr{C}.$ In particular, the third equality in Eq.~(85) can be useful from an experimental
viewpoint given that a small toroidal magnet (such as that employed by Tonomura et al. \cite{11,12,13}) may be modelled by a closed flux line and the topological invariance $L \barlambda \delta(C,\mathscr{C})/d=\Delta_{AB}=L \barlambda \delta(C,\mathscr{C}')/d$ implies that any possible deviation from a circular toroidal magnet in an experiment is insensitive to the AB shift.

The invariance $\delta(C, \mathscr{C})= \delta(\mathscr{C},C)$ is physically interesting. The basis of this invariance is the relation $l(C, \mathscr{C})=l(\mathscr{C},C)$ which is well-known in knot theory \cite{41} and was discovered by Gauss in 1833 \cite{37}! As already noted, this Gaussian relation admits a simple mathematical interpretation: the linking number $l$ is invariant under the simultaneous interchange of $C$ and $\mathscr{C},$ i.e. $C\to \mathscr{C}$ and $\mathscr{C} \to C$. It is clear that $\delta(C, \mathscr{C})$ is the AB phase that accumulates the wave function of a charge $q$ when it travels the path $C$ around the closed flux line of shape $\mathscr{C}.$ Therefore the relation $\delta(C, \mathscr{C})= \delta(\mathscr{C},C)$ tells us that $\delta(\mathscr{C},C)$
is the AB phase that accumulates the wave function of a charge $q$ when it travels the path $\mathscr{C}$ around the closed flux line of shape $C$.
This means that the charge $q$ and the steady current $\v J$ interchange their original curves. However, we must say that this interchange
is rather a mathematical procedure than a physical one ---it is assumed that the charge $q$ first travels on the curve $C$ and after on the curve $\mathscr{C}$ and that the current $\v J$ flows first on the curve $\mathscr{C}$ and after on the curve $C$. This requires that the charge $q$ and the current $\v J$ are free to travel different curves, i.e. that the charge $q$ and the current $\v J$ are not rigidly ``endowed'' to their initial curves.

\section{\large $\bfA$-explanation vs $\bfB$-explanation: local and nonlocal interpretations of the AB effect}
\label{7}
Although throughout this paper we have not hesitated to interpret the AB phase as a consequence of the nonlocal action of the magnetic field on the moving charge, we must recognize that this is not the most popular interpretation of the AB phase which considers either an infinitely-long solenoid or an infinitely-long flux line. In connection with this point we must say that there has been a longstanding debate between those who argue that the AB effect is caused by the local action of the vector potential on the moving charge (the $\bfA$-explanation), in which case this potential must be considered as a real physical quantity despite its gauge dependence, and those who argue that the AB effect is caused by the nonlocal action of the magnetic field on the moving charge (the $\bfB$-explanation), in whose case a form of action-at-a-distance is physically feasible (a short review of this debate can be found in Eynck et al. \cite{42}). This debate is currently unsettled. Many papers have addressed this debate \cite{43,44,45,46,47,48,49,50,51,52} and proposed other models that intend to explain the AB effect \cite{53,54,55,56,57,58,59,60,61,62,63,64,65,66,67,68,69,70,71,72,73,74,75,76}. This debate may be naturally translated to the case of the AB effect involving a vector potential outside a closed flux line and a magnetic field confined along this closed flux line. In this section we will address the $\bfA$-explanation and the $\bfB$-explanation of the AB phase in the context of the closed flux line.

We have seen that the AB phase can be expressed as $\delta =[q/(\hbar c)]\oint_C \v A \cdot d\v x$. If we apply the Stokes theorem in the considered  non-simply connected region: $\oint_{C>\partial S_{\varepsilon}} \v A \cdot d\v x= \int_{S_{\varepsilon}} \v B \cdot d \v S$ then the AB phase can be expressed as $\delta = [q/(\hbar c)]\int_{S_{\varepsilon}} \v B \cdot d \v S$. Combining both expressions for the AB phase we have
\begin{equation}
\frac{q}{\hbar c}\oint_{C>\partial S_{\varepsilon}} \v A \cdot d \v x=\delta=\frac{q}{\hbar c}\int_{S_{\varepsilon}} \v B \cdot d \v S.
\end{equation}
The $\bfA$-explanation is supported by the first equality in Eq.~(86). This states that the vector potential locally acts through its circulation on the charged particle originating the AB phase $\delta$ and therefore it makes sense to claim that the vector potential locally influences the phase of the wave function of the charged particle. Argued differently, the AB phase must be physically originated by the local action of an electromagnetic quantity in the region outside the closed flux line and since the only electromagnetic quantity defined in each point of that region is the vector potential $\v A$ then this potential should produce the AB phase $\delta$. In short: the vector potential $\v A$ exists in each point of the trajectory of the charge $q$ and therefore $\v A$ locally acts on $q$ producing $\delta$. On the other hand, the $\bfB$-explanation is supported by the second equality in Eq.~(86), which states that the magnetic field nonlocally acts through its flux on the charged particle originating the AB phase $\delta$. In short: $\v B$ exists along the closed flux line and not in each point of the trajectory of the charge $q$ outside this closed flux line and therefore $\v B$ nonlocally acts on $q$ producing $\delta$.

In this paper we have supported the $\bfB$-explanation and rejected the $\bfA$-explanation. Let us point out three arguments that strengthen our support for the former explanation:
\vskip 3pt
\noindent \textbf{(I)} The vector potential $\v A$ is gauge-dependent and therefore has no physical meaning.
\vskip 3pt
\noindent \textbf{(II)} The Stokes theorem applied in the non-simply connected region discussed in this paper (see Eq.~(41)) implies the relation
\begin{equation}
\frac{q}{\hbar c}\oint_{C_k>\partial S_{\varepsilon}}\!\v A\cdot d \v x=...=\frac{q}{\hbar c}\oint_{C_2>\partial S_{\varepsilon}}\! \v A \cdot d \v x=\frac{q}{\hbar c}\oint_{C_1>\partial S_{\varepsilon}}\! \v A \cdot d \v x=\delta=\frac{q}{\hbar c}\int_{S_{\varepsilon}} \v B \cdot d \v S,
\end{equation}
where the different charge paths $C_1>\partial S_{\varepsilon}, C_2>\partial S_{\varepsilon}, ..., C_k>\partial S_{\varepsilon}$ possessing all of them the same linking number are homotopically equivalent. If we consider the equalities on the left-hand side of $\delta$ in Eq.~(86) then there is a manifest ambiguity because we cannot distinguish if this phase is connected with the circulation of the vector potential $\v A$ along $C_1>\partial S_{\varepsilon}$ or $C_2>\partial S_{\varepsilon},$ or along $C_k>\partial S_{\varepsilon}$. These circulations are spatially delocalised with respect to the closed flux line (they are not functions of point) and therefore we cannot know which of the charge paths is locally connected with the circulation of the vector potential. In short: the vector potential $\v A$ is ambiguous due to its gauge-dependence and its circulation $\oint_{C}\v A\cdot d \v x$ is gauge invariant but it is ambiguous due to its spatial delocalisation (indistinguishability of the curve $C$). In consequence the $\bfA$-explanation does not hold because the alleged local action of the vector potential is ambiguous. But if we consider the last equality in Eq.~(86) then we conclude that the AB phase evaluated outside the closed line of singularity $\mathscr{C}$ is unambiguously connected with the confined magnetic flux along this closed line of singularity. Since the charge and the closed flux line lie in different spatial regions this magnetic flux nonlocally acts on the charge producing the AB phase. Thus, the $\bfB$-explanation holds. Let us remark that to understand the AB effect, we should be clear about the difference between applying the Stokes theorem in a simply connected region than in a non-simply connected region. This is the key to see why the AB phase is a nonlocal phase.

\vskip 3pt
\noindent \textbf{(III)} Equation~(45) implies the relation
\begin{equation}
\frac{q}{\hbar c}\int_{\Gamma}\v A\cdot d \v x'-\frac{q}{\hbar c}\int_{\gamma}\v A\cdot d \v x'= \frac{q}{\hbar c}\oint_{C}\v A\cdot d \v x',
\end{equation}
which can be written as
\begin{equation}
\frac{q}{\hbar c}\int^{\v x}_{{\scriptsize \bfcalO[\Gamma]}}\v A\cdot d \v x'+\frac{q}{\hbar c}\int_{\v x}^{{\scriptsize \bfcalO[\gamma]}}\v A\cdot d \v x'= \frac{q}{\hbar c}\oint_{C}\v A\cdot d \v x',
\end{equation}
where $[\Gamma]$ and $[\gamma]$ denote different paths going from the point $\small\bfcalO$ to the point $\v x$ (see Fig.~\ref{7}).
\begin{figure}
	\centering
	\includegraphics[scale=0.5]{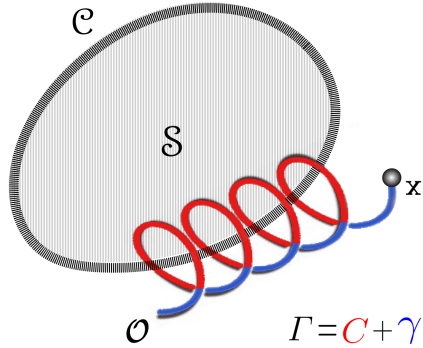}
	\caption{\small {The charged path $\Gamma$ can be decomposed as the sum of closed paths $C$ (red line) enclosing the closed flux line plus the open path $\gamma$ (blue line) that remains after the charge takes another turn around the solenoid. Both paths $\Gamma$ and $\gamma$ start from $\bfcalO$ to the variable point $\v x$. The path $\Gamma$ encloses the closed flux line while the path $\gamma$ does not.}}
	\label{Fig7}
\end{figure}
Note that
\begin{equation}
\oint_{C}\v A\cdot d \v x'=\int_{{\scriptsize\bfcalO[\Gamma]}}^{{\scriptsize\bfcalO[\gamma]}}\v A\cdot d \v x'\not=0,
\end{equation}
because the function $\Omega$ in the potential $\v A=\Phi\nabla\Omega/(4\pi)$ is a multi-valued function. Moreover, since $\v x$ is a variable point it follows that the quantities on the left-hand side of Eq.~(89) are local functions which may be defined as
\begin{equation}
\frac{q}{\hbar c}\int^{\v x}_{{\scriptsize\bfcalO[\Gamma]}}\v A\cdot d \v x'=\delta_{\Gamma}(\v x),\quad\frac{q}{\hbar c}\int_{\v x}^{{\scriptsize\bfcalO[\gamma]}}\v A\cdot d \v x'=\delta_{\gamma}(\v x).
\end{equation}
Considering these definitions of the functions $\delta_{\Gamma}(\v x)$ and $\delta_{\gamma}(\v x)$, it follows that the AB phase $\delta=[q/(\hbar c)]\oint_{C}\v A\cdot d \v x'$ in Eq.~(89) can be expressed as
\begin{equation}
\delta_{\Gamma}(\v x) + \delta_{\gamma}(\v x) =\delta,
\end{equation}
and thus one would be tempted to say that the AB phase is locally generated, a statement that supports the $\bfA$-explanation of the AB effect. Nevertheless, the fundamental problem with this argument is that both functions $\delta_{\Gamma}(\v x)$ and  $\delta_{\gamma}(\v x)$ are not gauge invariant. In fact, by applying the gauge transformation $\v A'=\v A+\nabla\Lambda$ to the functions in Eq.~(91) we have
\begin{equation}
\frac{q}{\hbar c}\int^{\v x}_{{\scriptsize\bfcalO[\Gamma]}}\v A'\cdot d \v x'=\frac{q}{\hbar c}\int^{\v x}_{{\scriptsize\bfcalO[\Gamma]}}\v A\cdot d \v x'+\frac{q}{\hbar c}[\Lambda(\v x)-\Lambda({\small\bfcalO})],
\end{equation}
\begin{equation}
\frac{q}{\hbar c}\int_{\v x}^{{\scriptsize\bfcalO[\gamma]}}\v A'\cdot d \v x'=\frac{q}{\hbar c}\int_{\v x}^{{\scriptsize\bfcalO[\gamma]}}\v A\cdot d \v x'+\frac{q}{\hbar c}[\Lambda({\small\bfcalO})-\Lambda(\v x)].
\end{equation}
These equations show that both $\delta_{\Gamma}(\v x)$ and  $\delta_{\gamma}(\v x)$ are not gauge invariant and therefore they cannot be considered the physical causes of the AB phase, i.e. these functions are not measurable quantities and consequently they are devoid of any physical meaning. Only the sum of these local functions is physically meaningful because it is gauge invariant ---this sum is no longer a function of point but a constant quantity implying the vanishing of the locality of the AB phase. Accordingly, on the basis of the arguments (I)-(III) we admit the $\bfB$-explanation and reject the $\bfA$-explanation.

As already pointed out, the nonlocal interpretation of the AB effect is much less popular than its local interpretation. One of the supporters of the former interpretation is Aharonov who in 1983 claimed \cite{77}: \emph{``In this talk we will review the AB effect, which .... provides a particularly clear example of nonlocal phenomena...One hopes that the arguments in the preceding section have convinced the reader that the A-B effect is indeed nonlocal.''} Here Aharonov clearly adopted the nonlocal explanation of the AB effect, and in doing so he disregarded the local explanation he had initially held together with Bohm \cite{1}. More recently, Aharonov et al. have pointed out \cite{78}: \emph{``The {\rm [AB]} phase is topological because it is determined by the number of windings the charge carries out around the solenoid, and is independent of the details of the trajectory. The phase is also nonlocal: while the magnetic flux in the solenoid clearly affects the resulting interference pattern, it has no local observable consequences along any point on the trajectory.''} Aharonov and Rohrlich have also pointed out \cite{2}: \emph{``Thus, instead of concluding that $\v A$ and $V$ are physical variables in quantum mechanics, we state a conclusion ... : Only $\v E$ and $\v B$ are physical quantities, but they act nonlocally ---a magnetic field here has physical effects on electrons there, and so on. Such action at a distance by a field is completely nonclassical.''}

The idea that the AB effect is purely quantum has been emphasised by many authors and experiments confirming such an idea have been carried out \cite{79,80,81}. Considering that the AB effect is a quantum-mechanical effect one could come to the conclusion that the nonlocality of the magnetic field in this effect is also inherently quantum-mechanical, as pointed out by Aharonov and Rohrlich \cite{2}. However, we have emphasised here that the nonlocality of the AB effect is of topological nature and therefore one could reasonably expect that this nonlocality could arise in other branches of physics ---after all topology is a branch of mathematics independent of quantum mechanics. With regard to this point, it is pertinent to say that in two recent papers \cite{38,39} we have discussed three classical electromagnetic configurations defined in non-simply connected regions: (i) an electric charge encircling an infinitely-long solenoid, (ii) a magnetic charge encircling an infinitely-long electric solenoid, and (iii) a dyon encircling an infinitely-long dual solenoid enclosing magnetic and electric fluxes. We have shown that the electromagnetic angular momenta arising from these configurations describe nonlocal interactions between the encircling charges (electric, magnetic, or dyon) outside the respective solenoids and the corresponding fluxes (magnetic, electric, or dual) confined inside these solenoids. In particular, we have argued \cite{38} that the electromagnetic angular momentum of the configuration (i) may be considered as the classical counterpart of the AB effect. It is pertinent to note here that a number of authors have proposed classical analogues of the AB effect \cite{78,82,83,84,85,86,87,88,89,90}. We think that the AB effect is a purely quantum effect but the nonlocality of this effect also appears in classical effects. Anyway, the fact that this nonlocal feature arises in both classical and quantum physics illustrates the power of topology in physics. As already noted: \emph{topology dictates nonlocality!}

\section{\large A gauge that eliminates the vector potential in all space except in a finite region}
\label{8}
In the previous section we have emphasised three formal arguments against the local $\bfA$-explanation of the AB effect. In this section we will present a fourth formal argument, which may be even stronger and convincing than the first three.

Consider the vector potential of the closed flux line $\v A =\Phi\nabla\Omega/(4\pi)=\Phi\nabla\Omega_0/(4\pi)+ \Phi \bm{\delta}_{\mathscr{S}}$. This potential satisfies $\nabla \times \v A = \Phi \bm{\delta}_{\mathscr{C}}=\v B$. Let us write the expression of this potential as
\begin{equation}
\v A +\nabla\Lambda= \Phi \bm{\delta}_{\mathscr{S}},
\end{equation}
where $\Lambda$ is a scalar function defined as
\begin{equation}
\Lambda=-\frac{\Phi}{4\pi}\Omega_0.
\end{equation}
A crucial observation is that the function $\Lambda$ is single-valued and this means that this function satisfies the Schwarz integrability condition in all space $(\partial^i \partial^j- \partial^j\partial^i)\Lambda=0,$ a condition that follows from the single-valuedness of the solid angle function $\Omega_0$ which satisfies $(\partial^i \partial^j- \partial^j\partial^i)\Omega_0=0$. This is explicitly demonstrated Appendix C. Considering Eq.~(95) we may interpret the function $\Lambda$ as the single-valued gauge function of the non-singular gauge transformation
\begin{equation}
\v A'=\v A +\nabla\Lambda,
\end{equation}
that transforms the Coulomb-gauge potential $\v A= \Phi\nabla\Omega/(4\pi)$ into the potential
\begin{equation}
\v A'=\Phi \bm{\delta}_{\mathscr{S}}.
\end{equation}
Remarkably, the gauge function $\Lambda$ in Eq.~(96) has the unusual property of changing the domain of definition of the potential $\v A$, i.e. while the potential $\v A$ is defined in all space outside the closed flux line, the potential $\v A'$ vanishes in all space except on the surface $\mathscr{S}$ surrounded by the closed flux line. To see this peculiar behaviour more clearly, we can use Eq.~(10) in Eq.~(98) to obtain
\begin{equation}
\v A'=\Phi \int_{\mathscr{S}} \delta(\v x- \v x')\,d \v S',
\end{equation}
which shows that for any point $\v x$ not on the surface $\mathscr{S}$ the potential $\v A'$ vanishes, i.e. $\v A'(\v x \notin \mathscr{S})=0.$ Evidently, $\nabla \times \v A' =\nabla \times \v A=\v B$ because $\nabla \times \nabla\Lambda= 0$ ($\Lambda$ is single-valued). While the vector potential $\v A$ satisfies the Coulomb gauge condition $\nabla \cdot \v A=0,$ the vector potential $\v A'$ satisfies the gauge condition
\begin{equation}
\nabla\cdot\v A'= \Phi \nabla\cdot\bm{\delta}_{\mathscr{S}},
\end{equation}
or equivalently
\begin{equation}
\nabla\cdot\v A'= -\Phi\int_{\mathscr{S}}\nabla'\delta(\v x- \v x')\cdot d \v S',
\end{equation}
where we have taken the divergence to Eq.~(99) and used $\nabla \cdot [\delta(\v x- \v x')d\v S']= \nabla\delta(\v x- \v x') \cdot d \v S'$ and $\nabla\delta(\v x- \v x')=-\nabla'\delta(\v x- \v x').$ Since $\v A$ and $\v A'$ are equivalent potentials it follows that the AB phase calculated with the transformed potential $\v A'$ is the same as that calculated with the original potential $\v A$. This may be seen by using Eq.~(98):
\begin{equation}
\frac{q}{\hbar c} \oint_{C}\v A' \cdot d \v x = \frac{q \Phi}{\hbar c}\oint_{C} \bm{\delta}_{\mathscr{S}}\cdot d \v x= l\frac{q\Phi}{\hbar c},
\end{equation}
where we have used the definition of the linking number given in Eq.~(27). Accordingly, the equivalence $[q/(\hbar c)]\oint_{C}\v A' \cdot d \v x =\delta= [q/(\hbar c)]\oint_{C}\v A \cdot d \v x$ holds and thereby those gauge-invariant quantities discussed in this paper involving the Coulomb-gauge potential $\v A$, for example the two-slit interference shift $\Delta_{AB}=L\barlambda q\Phi/d$, are identical if we use the potential $\v A'$ in the new gauge specified by Eq.~(101). We should also note that Eq.~(102) verifies our earlier assertion made in Eq.~(51) that the AB phase in a closed flux line is not continuously accumulated along the charge path $C$ but discretely picked up as the charge crosses the closed flux line.

Interestingly, Eq.~(98) can be expressed as $[\Phi/(4\pi)]\nabla(\Omega - \Omega_0)=\v A'$ which shows that the vector potential in the introduced gauge may be written as
\begin{equation}
\v A' = \frac{\Phi}{4\pi}\nabla \Theta,
\end{equation}
where $\Theta = \Omega- \Omega_0$ is the function formed by the difference between the multi-valued representation of the solid angle $\Omega$ and the single-valued representation of the solid angle $\Omega_0.$ Since $\Omega$ is a multi-valued function it follows that $\Theta$ is a multi-valued function and therefore $(\partial^i \partial^j- \partial^j\partial^i)\Theta\neq 0.$
\begin{figure}
	\centering
	\includegraphics[scale=0.60]{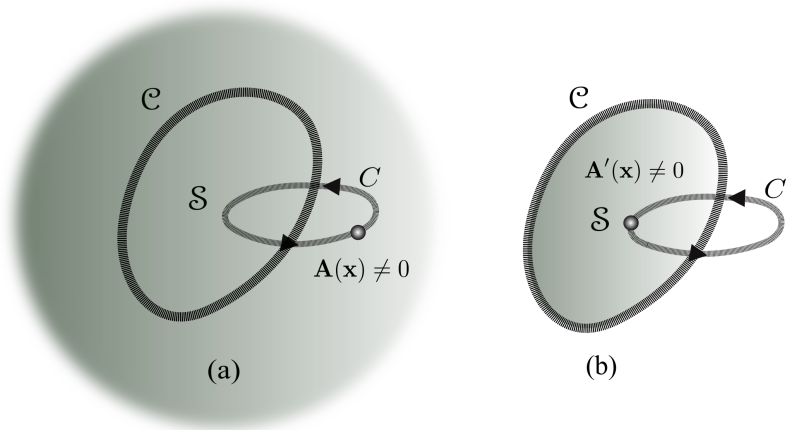}
	\caption{\small {(a) The potential $\v A(\v x)$ is non-vanishing in every point $\v x$ along the path $C$ of the charged particle. (b) The  potential $\v A'(\v x)$ is zero in every point $\v x$ along the path $C$ of the charged particle except at a single point localised on the surface $\mathscr{S}.$}}
	\label{Fig8}
\end{figure}

According to the $\bfA$-explanation of the AB effect, the vector potential exists in every point of the trajectory of the charged particle and
then quantum mechanics may be invoked to argue that this potential, through its  gauge invariant circulation, influences the wave function of this particle producing the AB phase [see Fig.~\ref{Fig8} (a)]. However, we have introduced here a non-singular gauge in which the vector potential [Eq.~(99)] vanishes in all space except on the surface surrounded by the closed flux line  implying that every time the charge takes a turn around the closed flux line the vector potential is zero along the trajectory of the charge expect on a single point of this trajectory  ---the crossing point where the trajectory of the charge intersects the surface surrounded by the closed flux line [see Fig.~\ref{Fig8} (b)]. In short: the vector potential exists only at a point of the charge path. The widely spread idea that the vector potential  ``acts'' through its circulation on the charge at every point of its path then collapses.

At the risk of being too reiterative, let us insist in our fourth argument against the $\bfA$-explanation. As is well-known, the $\bfA$-explanation of the AB effect tells us that the original vector potential $\v A$ exists at every point of the trajectory of the charge. But we have showed here that the transformed vector potential $\v A'$ in Eq.~(99) is zero along this trajectory except on one of its points. Since we can always transform $\v A$ into $\v A'$ via a non-singular gauge transformation [Eq.~(97)] then, after this transformation, the domain of definition of the vector potential has been modified. This behaviour of the vector potential does not correspond to any physical local quantity since the domain of definition of a physical quantity (in a reference frame) should remain fixed in order for the local interaction of this quantity to make sense. In this regard, it is interesting to note that Paiva et al. \cite{91} have recently pointed out a similar argument in favour of a nonlocal interpretation of the AB effect (they considered an infinitely-long flux line): \textit{``...because there always exists a gauge in which vector potential vanishes in an arbitrary region that does not completely enclose the solenoid, a priori, the AB effect cannot be seen as the result of the local interaction between the charge and the vector potential.''} However, these authors did not provide any explicit gauge that proves their statement. Here we have explicitly shown that such a gauge exists in the AB effect in a closed flux line (the corresponding gauge condition is given in Eq.~(101)). In a few words: the original potential $\v A$ exists at every point on
the charge path but the transformed potential $\v A'$ exists at a single point of the charge path! The alleged local interaction of this vector potential with the particle could be anything but a physical interaction.
\section{\large Singular and non-singular gauge transformations in the AB effect}
\label{9}
The term ``singular gauge transformations'' deserves a clarification. In proper physical jargon, gauge transformations
are transformations of potentials that leave invariant their associated fields. End of story. But physicists are sloppy in the use of language and some of them have wanted to extend the story by talking about ``singular'' gauge transformations. In fact, the term singular gauge transformations is an oxymoron because these transformations are not true gauge transformations ---they modify their corresponding fields! If we admit such singular gauge transformations then we could appear or disappear magnetic fields depending on the chosen gauge. In Kleinert's words \cite{33}: \textit{ ``...Obviously, this terminology} [singular gauge transformation] \textit{is misleading and must be rejected. After all, if we were to allow for such ``singular” (i.e. nonintegrable) transformations...we could reach an arbitrary field $F_{\mu \nu}$ starting from $F_{\mu \nu}\equiv 0$, and the physics would certainly not be invariant under this.''} The failure to specify the difference that exists between applying either a singular or non-singular gauge transformations in the context of the AB effect may lead to the misleading conclusion that the vector potential in the AB effect can be eliminated via a gauge transformation.

Consider the following statement: unlike a non-singular gauge, a singular gauge modifies the magnetic field and therefore it cannot be considered a symmetry of this field. This assertion should be included in textbooks and then one could understand that the possible disappearance of  the vector potential in a given configuration via a singular gauge entails a modification of the magnetic field ---which totally breaks the spirit of gauge invariance. The puzzling question arises: how to understand a supposed gauge transformation that changes the magnetic field?

In order to see how a singular gauge was introduced to interpret the AB effect, let us do some history. In 1979 Bocchieri and Loinger \cite{92} surprised theoretical physicists by claiming that they had found a gauge in which the vector potential is zero outside the infinitely-long solenoid and as a consequence the AB effect disappeared in this gauge! These authors then concluded that the AB effect \cite{92}:\emph{``...has a purely mathematical origin''}, i.e. this effect was a sort of mathematical artefact devoid of physical meaning. The argument of Bocchieri and Loinger can be reconstructed as follows.

Consider first the vector potential of an infinitely-long solenoid of radius $R$.  In the Coulomb gauge  this vector potential can be written in cylindrical coordinates as \cite{38,93}
\begin{equation}
\v A = \frac{\Phi}{2\pi}\bigg[\frac{\Theta(\rho-R)}{\rho} + \frac{\rho\,\Theta(R-\rho)}{R^2}  \bigg]\,\hat{\!\bfphi},
\end{equation}
where $\Phi =\pi R^2 B$ is the flux through the solenoid with $B$ being the magnitude of the magnetic field inside the solenoid and $\Theta$ is the Heaviside step function. The curl of Eq.~(104) gives \cite{38,93}
\begin{equation}
\nabla \times \v A = \frac{\Phi}{\pi R^2}\Theta(R-\rho)\hat{\v z},
\end{equation}
whose right-hand side identifies with  the magnetic field confined in the solenoid \cite{38}: $\v B= \Phi\Theta(R-\rho)\hat{\v z}/(\pi R^2)$.
Following Bocchieri and Loinger \cite{92}, let us apply a ``presumable'' gauge transformation to transform the potential $\v A$ into the potential $\v A'$ by adding to it the gradient of the function $\chi$
\begin{equation}
\v A' = \frac{\Phi}{2\pi}\bigg[\frac{\Theta(\rho-R)}{\rho} + \frac{\rho\,\Theta(R-\rho)}{R^2}  \bigg]\,\hat{\!\bfphi} +\nabla\chi,
\end{equation}
where $\chi=-\Phi\phi/(2\pi)$ with $\phi$ being the azimuthal angle.

By invoking the usual result that the curl of a gradient identically vanishes, it follows that
\begin{equation}
\nabla \times \v A' = \frac{\Phi}{\pi R^2}\Theta(R-\rho)\hat{\v z},
\end{equation}
and then $\v A$ and $\v A'$ are equivalent potentials in the sense that the curl of both yields the same magnetic field.

Now $\nabla\chi=-\Phi\,\hat{\!\bfphi}/(2\pi \rho)$ and therefore Eq.~(106) can be written as
\begin{equation}
\v A' = \frac{\Phi}{2\pi\rho}\bigg[\Theta(\rho-R)-1\bigg]\,\hat{\!\bfphi} +  \frac{\Phi}{2\pi}\frac{\rho\,\Theta(R-\rho)}{R^2}\,\hat{\!\bfphi}.
\end{equation}
Outside the solenoid $\textbf{[}\rho>R, \Theta(\rho-R)\!=\!1$ and $\Theta(R-\rho)\!=\!0\textbf{]}$  the potential vanishes: $\v A'_{\text{out}}=0$ while inside the solenoid $\textbf{[}R>\rho, \Theta(R-\rho)\!=\!1$ and $\Theta(\rho-R)\!=\!0$\textbf{]} the potential takes the form $\v A'_{\text{in}}= [\rho\Phi/(2\pi R^2)-\Phi/(2\pi\rho)]\,\hat{\!\bfphi}$. Since the charge path $C$ is outside the solenoid then it follows that $\delta=[q/\hbar c]\oint_{C}\v A'_{\text{out}}\cdot d \v x=0$ and therefore the AB effect vanishes. The conclusion arises: the AB phase exists for the potential $\v A$ but not for the equivalent potential $\v A'$. This means that the AB phase is gauge dependent and therefore the AB effect seems to be a mathematical construct devoid of physical meaning. As Bocchieri and Loinger claimed \cite{92}: \emph{``...Obviously, in this gauge ${\bm[}\v A'_{\rm {out}}=0{\bm]}$ there is no Aharonov-Bohm effect. This way of reasoning shows that the effect is gauge-dependent; this seems astonishing, at first,
as in ${\bm [}\v A_{\rm {out}}=\Phi\,\hat{\!\bfphi}/(2\pi\rho){\bm]}$ only $\Phi$ appears, which is a gauge-invariant concept.''}

The argument of Bocchieri and Loinger seems to be correct except for one thing: the function $\chi=-\Phi\phi/(2\pi)$ is a multi-valued function and its gradient $\nabla\chi=-\Phi\,\hat{\!\bfphi}/(2\pi \rho)$ is a singular function. As already noted, if $\Lambda$ is a single-valued function then it satisfies the
Schwarz integrability condition: $(\partial^i \partial^j- \partial^j \partial^i)\Lambda= 0$. The single-valuedness of $\Lambda$ implies $\nabla \times \nabla\Lambda =0$. But if $\Lambda$ is a multi-valuated function then it violates the Schwarz condition: $(\partial^i \partial^j- \partial^j \partial^i)\Lambda \neq 0$ and therefore $\nabla \times \nabla\Lambda \neq 0$.  The function considered by Bocchieri and Loinger:   $\chi=-\Phi\phi/(2\pi)$ is a multi-valued function that satisfies
\begin{equation}
\nabla \times \nabla \chi = -\frac{\Phi}{2\pi}\frac{\delta(\rho)}{\rho}\hat{\v z},
\end{equation}
which can be verified as follows. From the Stokes theorem we can write $\oint_{C}\nabla\chi\cdot d \v x = \int_{S}\nabla \times \nabla \chi \cdot d \v S,$ where $C$ is the boundary of $S$ and it is assumed $C$ encircles the solenoid. Using $\chi = - \Phi \phi /(2\pi)$ we obtain $\oint_{C}\nabla\chi\cdot d \v x = [-\Phi/(2\pi)]\oint_{C}\nabla \phi \cdot d \v x$. Inserting $\nabla\phi=\hat{\bfphi}/\rho$ and $d \v x = d\rho \hat{\bfrho} + \rho d\phi \hat{\bfphi} + dz\hat{\v z}$ it follows $\oint_{C}\nabla \phi \cdot d \v x= \oint_{C}d\phi$ and since $\oint_{C}d\phi=2\pi n$ where $n$ is the winding number of the path $C$, it follows that $\oint_{C}\nabla\chi\cdot d \v x =-n\Phi$ which together with the Stokes theorem imply
the relation $\int_{S}\nabla\times \nabla\chi\cdot d \v S=-n\Phi$.  This relation is satisfied by Eq.~(109) which verifies the validity of this equation. Therefore the correct expression for the curl of the potential $\v A'$ in Eq.~(108) reads
\begin{equation}
\nabla \times \v A' = \frac{\Phi}{\pi R^2}\Theta(R-\rho)\hat{\v z}-\frac{\Phi}{2\pi}\frac{\delta(\rho)}{\rho}\hat{\v z}.
\end{equation}
The first term on the right-hand side identifies with the  magnetic field confined in the solenoid $\v B= \Phi\Theta(R-\rho)\hat{\v z}/(\pi R^2)$
while the second term identifies with the negative of the singular magnetic field $\v B_\texttt{string}=\Phi \delta(\rho)\hat{\v z}/(2\pi\rho)$
due to an infinitely-long flux line (or a magnetised string) localised along the $z-$axis. Expressed more compactly Eq.~(110) reads
\begin{equation}
\v B' = \v B-\v B_\texttt{string}.
\end{equation}
It becomes now evident that the argument of Bocchieri and Loinger for the elimination of the AB effect is inconsistent because
it breaks the gauge invariance of the magnetic field of the solenoid: $\v B=\nabla \times \v A\neq\v B'=\nabla \times \v A'$ ---several authors \cite{93,94,95,96} have noted this inconsistence of the supposed gauge transformation proposed by Bocchieri and Loinger.

The gauge transformation applied by Bocchieri and Loinger is an example of the so-called singular gauge transformations. The wrong idea that singular gauge transformations do not modify the magnetic field has been something recurrent in the literature. For example, Wilczek \cite{97} followed an argument similar to that of Bocchieri and Loinger to eliminate the vector potential outside an infinitely-long solenoid by means of a non-singular gauge transformation in order to obtain the eigenvalues of the $z-$component of the angular momentum of the system formed by a charge encircling this solenoid. The inconsistency of Wilczek's approach was pointed out by Kobe \cite{98}. Nambu \cite{99} also used a singular gauge transformation to study the AB effect. Other authors have also used singular gauge transformations to study the quantum mechanics of a charge encircling an infinitely-long solenoid (see, for example, Ref.~\cite{100}).

Similar arguments to those of Bocchieri and Loinger can be applied to the vector potential of the closed flux line and its associated AB effect.
In the previous section we have applied the non-singular gauge transformation
$\v A'=\v A +\nabla\Lambda$ with $\Lambda=-\Phi\Omega_0/(4\pi)$ to transform the Coulomb-gauge potential $\v A= \Phi\nabla\Omega/(4\pi)$ into the potential
$\v A'=\Phi \bm{\delta}_{\mathscr{S}}.$ Suppose now that instead of considering the gauge function $\Lambda=-\Phi\Omega_0/(4\pi)$ we consider the ``gauge'' function
$\chi=-\Phi\Omega/(4\pi)$. It follows that
\begin{equation}
\v A'=\frac{\Phi}{4\pi}\nabla\Omega+\nabla\chi=\frac{\Phi}{4\pi}\nabla\Omega-\frac{\Phi}{4\pi}\nabla\Omega=0.
\end{equation}
Apparently, we have found a gauge in which the vector potential is zero and therefore the AB effect in the closed flux line is seen to disappear! Nevertheless, we had to pay a very high price: while the curl of the original potential $\v A$ in Eq.~(16) yields the non-vanishing magnetic field $\v B= \Phi\oint_{\mathscr{C}}\delta(\v x- \v x')\, d \v x'$, the curl of the transformed potential $\v A'=0$ yields a vanishing magnetic field $\v B'=0$. The transformation we have applied here is a singular gauge transformation which modifies the magnetic field ---the function $\chi=-\Phi\Omega/(4\pi)$ is a multi-valued function.

An alternative way to see the difference between non-singular and singular gauge transformations may be given  using index notation in which
the magnetic field is represented by the axial component $F^{ij}=-\varepsilon^{ijk}B_k $ (where $B_k$ represents the components of the field $\v B$) of the electromagnetic field tensor $F^{\mu\nu}$ (for notation and conventions in the Minkowski spacetime see Ref.~\cite{101}). In our case this axial component [the magnetic field in Eq.~(4)] reads
\begin{equation}
F^{ij}= -\Phi \varepsilon^{ijk}(\bm{\delta}_{\mathscr{C}})_k,
\end{equation}
From $F^{\mu \nu}=\partial^{\mu}A^{\nu}-\partial^{\nu}A^{\mu}$ and $\partial^{\mu}=(\partial^{0},-\partial^i)$ it follows that the tensor $F^{ij}$ can be expressed in terms of the components $A^j$ of the vector potential $\v A$ as
\begin{equation}
F^{ij}=\partial^j A^i - \partial^i A^j,
\end{equation}
and in terms of  the primed components  $A'^j$ of the primed vector potential $\v A'$ by $F'^{ij}=\partial^j A'^i - \partial^i A'^j$. Let us apply the transformation  $A^j \to A^j=A'^j-\partial^j\Lambda$, where $\Lambda$ is an arbitrary function of space (at this stage this transformation is not necessarily a gauge transformation), to Eq.~(114). We obtain
\begin{equation}
F^{ij}=F'^{ij} +(\partial^i \partial^j - \partial^j \partial^i )\Lambda.
\end{equation}
If $\Lambda$ is a single-valued function then it satisfies  the Schwarz condition $(\partial^i \partial^j - \partial^j \partial^i )\Lambda=0$. In this case the corresponding potentials are gauge potentials and their associated transformation is a gauge transformation which leaves invariant the axial component of the electromagnetic field: $F^{ij}=F'^{ij}$. But if $\Lambda$ is a multi-valued function then it does not satisfy  the Schwarz condition $(\partial^i \partial^j - \partial^j \partial^i )\Lambda\not=0$. In this case the corresponding potentials are not gauge potentials and its associated transformation is  not a gauge transformation because it  does not leave invariant the axial component of the electromagnetic field: $F^{ij}\not=F'^{ij}$. In our particular case,  we have the specific multi-valued gauge function  $\Lambda=-\Phi\Omega/(4\pi)$ which satisfies $(\partial^i \partial^j - \partial^j \partial^i )\Lambda=- \Phi \varepsilon^{ijk}(\bm{\delta}_{\mathscr{C}})_k$. When this result and Eq.~(113) are used in Eq.~(115) we conclude that $F'^{ij}=0$. In this case we have a singular gauge transformation that transforms the non-vanishing field $F^{ij}= -\Phi \varepsilon^{ijk}(\bm{\delta}_{\mathscr{C}})_k$ into the vanishing field $F'^{ij}=0!$ As Sidney Coleman said \cite{102}: \emph{``...singular gauge transformations make people uneasy.''}

\section{\large Concluding remarks}
\label{10}
\noindent Millikan once said  \cite{103}:
\emph{``Science walks forward on two feet, namely theory and experiment, ... Sometimes it is one foot that is put forward first, sometimes the other, but continuous progress is only made by the use of both —by theorizing and then testing, or by finding new relations in the process of experimenting and then bringing the theoretical foot up and pushing it on beyond, and so on in unending alterations.''} Millikan's quote applies very well to the case of the AB effect:
\vskip 5pt
\noindent \textbf{The first step was with the theoretical foot:}
\vskip 2pt
The story began in 1959 when Aharonov and Bohm \cite{1} theoretically predicted the now known as AB effect in which a relative phase shift proportional to the magnetic flux of an ``idealised'' infinitely-long solenoid can be observed as a displacement of interference fringes, even when the interfering charged particles pass through a spatial region where there is no magnetic field but there is a vector potential. Two problems did not take long to appear:
\vskip 3pt
\noindent (i) The effect was predicted considering an idealised infinitely-long solenoid which was evidently unavailable.
\vskip 3pt
\noindent (ii) The effect admitted two different and excluding interpretations: it was either attributed to the local action of the vector potential ($\bfA$-explanation) or to the nonlocal action of the magnetic field ($\bfB$-explanation).
\vskip 2pt
\noindent The solution to the problem (i) would be to replace the non-available infinitely-long solenoid with a long but finite solenoid which was attainable.  However, the finite long solenoid does not confine completely the magnetic field and therefore the use of this solenoid does not allow to theoretically predict the AB effect. No conclusive formal argument was proposed to find the correct explanation of the AB effect and thus to solve the problem (ii).
\vskip 5pt
\noindent \textbf{The second step was with the experimental foot:}
\vskip 2pt
In the 1960's several experiments that tried to test the AB effect were reported \cite{5,6,7,8} in which the corresponding displacement of interference fringes were observed using finite magnetic devices like magnetised whiskers and long solenoids. But these displacements were questioned \cite{9,10} due to the unavoidable magnetic field leakage rather than to the AB effect. However, in the 1980's Tonomura and collaborators \cite{11,12,13} decisively confirmed the AB effect in a series of experiments using a toroidal configuration. In a first instance \cite{11} they used a squared micro-sized toroidal magnet whose field leakage was sufficiently small to verify the AB effect with good accuracy. In a second instance \cite{12,13}, they used a circular micro-sized toroidal magnet covered with a superconducting layer which avoided the field leakage due to the Meissner effect. Thus, the experimental foot suggested a way to solve the problem (i) by detecting the AB effect using a tiny magnet whose shape was roughly that of a toroid. However, two questions arise:
\vskip 2pt
\noindent
(iii) Is there a well-established theoretical treatment of the AB effect using a toroidal configuration?
\vskip 2pt
\noindent
(iv) How to explain that the AB effect arises in two configurations of different geometry?
\vskip 2pt
\noindent Expectably, the theoretical foot could solve the problems (i)-(iv).
\vskip 5pt
\noindent  \textbf{The third step is with the theoretical foot:}
\vskip 2pt
We have addressed here the problem (i) by showing that the AB effect is theoretically predicted using a less-idealised closed flux line instead of using a highly-idealised infinitely long solenoid. With regard to the problem (ii) we have argued that the $\bfB$-explanation should be seen as the correct explanation of the AB effect. We have questioned the $\bfA$-explanation by introducing a non-singular gauge in which the vector potential vanishes in all space except on the surface surrounded by the closed flux line and therefore, as the charge encircles the closed flux line, the vector potential is zero along the trajectory of the charge except on a point of this trajectory, which questions its alleged physical significance. We have emphasised the difference in applying singular and non-singular gauge transformations in the AB effect and argue that only the latter are consistent with the gauge invariance of the magnetic field. With regard to the question (iii) we should first say that there have been some theoretical treatments on the AB effect in a toroidal solenoid \cite{14,15,17,104,105,106,107,108,109,110,111}. However these treatments are too cumbersome to the extent that most authors (if not all) of standard textbooks on quantum mechanics (see, for example, Weinberg \cite{112} and Sakurai \cite{113}) discussing theoretical aspects of the AB effect still prefer the use of an infinitely-long solenoid instead of a toroidal solenoid even though these authors justify the experimental validity of the AB effect with the experiments of Tonomura et al. \cite{11,12,13} who clearly used a toroidal configuration. Here we have addressed the question (iii) by showing that if a thin toroidal solenoid is modelled by a closed flux line of arbitrary shape and size then the theoretical treatment of the AB effect is exact and considerably accessible to the extent that it may be included in textbooks on quantum mechanics.  We have demonstrate that the AB phase in a closed flux line is determined by a linking number and exhibits the same form as the AB phase in an infinitely-long flux line which is determined by a winding number. We have showed that the shift in the interference pattern associated to the AB effect in a closed flux line is the same as that associated to an infinitely-long flux line. We have addressed the question (iv) by emphasising that the AB effect arises in different geometries because it is of topological nature. We have stressed the topological character of the AB phase in a closed flux line by introducing four topological invariances of this phase: invariance under deformations of the charge path, invariance under deformations of the closed flux line, invariance under simultaneous deformations of the charge path and the closed flux line, and invariance under the interchange between the charge path and the closed flux line.

Finally, we would like to comment on four further and feasible applications of the closed-flux-line model discussed here. The first application deals with the scattering amplitude of charges outside the closed flux line. Just like the AB scattering in an infinitely-long flux line, in an infinitely-long solenoid, or in a toroidal solenoid, we expect that the scattering amplitude of charged particles outside a closed flux line should have a non-vanishing effect due to the confined magnetic flux. The second application deals with the energy levels of a charged particle encircling the closed flux line. Similarly to the energy levels of the system formed by a charge encircling an infinitely-long flux line or an infinitely-long solenoid, we expect the energy levels of the system formed by a charged particle encircling the closed flux line to have an explicit dependence on the magnetic flux. The third application deals with demonstrating that the AB phase in a closed flux line is an example of the Berry phase \cite{114}. The fourth application deals with anyons or composite systems formed by a charge encircling a flux line which obey fractional statistics. Wilczek's \cite{115} original anyon model is based on a charge encircling an infinitely-long solenoid. The anyon model was later extended to charges encircling a toroidal solenoid \cite{116,117,118,119}, i.e. toroidal anyons. The quantum mechanics of closed-flux line anyons would be an interesting theoretical possibility.

\vskip 20pt


\noindent \textbf{Acknowledgments.} I thank my father Jos\'e A. Heras for the insightful and often enjoyable discussions we had about the AB effect. Perhaps the most
entertaining part of our conversations was trying to understand why
nonlocality seems to be such an unthinkable concept for many
physicists. As an undergraduate student, I wrote this paper in my
free time and with the interest of understanding the AB effect without any kind of prejudice.

\renewcommand\theequation{A\arabic{equation}}
\setcounter {equation}{0}
\section*{\large Appendix A. Derivation of Eq.~(4)}

Consider the potential of the closed flux line defined in Eq.~(7)
\begin{equation}
\v A=\frac{\Phi}{4\pi} \nabla\times \oint_{\mathscr{C}} \frac{d \v x'}{|\v x - \v x'|}.
\end{equation}
The curl of this potential and the use of the identity $\nabla \times (\nabla \times \v F) = \nabla (\nabla \cdot \v F)- \nabla^2 \v F$ yield
\begin{eqnarray}
\nonumber \nabla \times \v A=\frac{\Phi}{4\pi}\nabla \times \bigg[\nabla\times \bigg( \oint_{\mathscr{C}} \frac{d\v x'}{|\v x-\v x'|} \bigg)\bigg]\qquad\qquad\qquad\quad\,
\\
= \frac{\Phi}{4\pi} \bigg[\nabla \oint_{\mathscr{C}} \nabla \cdot \bigg( \frac{d\v x'}{|\v x-\v x'|} \bigg) -\oint_{\mathscr{C}}\nabla^2 \bigg(\frac{ d\v x'}{|\v x-\v x'|} \bigg)\bigg].
\end{eqnarray}
Inserting $\nabla \cdot (d \v x'/|\v x-\v x'|) = -\nabla'(1/|\v x\!-\!\v x'|) \cdot d\v x'$ and $\nabla^2( 1 / |\v x-\v x'|)=-4 \pi \delta(\v x - \v x')$ in Eq.~(A2), we obtain
\begin{equation}
\nabla \times \v A= \frac{\Phi}{4\pi} \bigg[-\nabla \oint_{\mathscr{C}} \nabla'  \bigg( \frac{1}{|\v x-\v x'|} \bigg)\cdot d\v x' +4\pi\oint_{\mathscr{C}}\delta(\v x- \v x')\,d\v x'\bigg].
\end{equation}
The first term on the right-hand side vanishes because $\oint_{\mathscr{C}}\nabla'(1/|\v x- \v x'|)\cdot d \v x'=0$ on account of the gradient theorem and the fact that $1/|\v x- \v x'|$ is a single-valued function of $\v x'$. Thus, Eq.~(A3) becomes
\begin{equation}
\nabla \times \v A= \Phi\oint_{\mathscr{C}}\delta(\v x- \v x')\,d\v x'=\v B,
\end{equation}
which shows that the curl of Eq.~(7) yields the magnetic field given in Eq.~(4).

\renewcommand\theequation{B\arabic{equation}}
\setcounter {equation}{0}
\section*{{\large Appendix B. Derivation of Eq.~(8)}}
Consider the potential of the closed flux line defined in Eq.~(7)
\begin{equation}
\v A=\frac{\Phi}{4\pi} \nabla\times \oint_{\mathscr{C}} \frac{d \v x'}{|\v x - \v x'|}.
\end{equation}
Using the Stokes theorem in the closed line integral of this potential we obtain
\begin{equation}
\oint_{\mathscr{C}}\frac{d \v x'}{|\v x- \v x'|}= \int_{\mathscr{S}}d\v S'\times\nabla'\bigg( \frac{1}{|\v x- \v x'|} \bigg),
\end{equation}
where $\mathscr{S}$ is the surface enclosed by the curve $\mathscr{C}.$ Making use of the relations $\nabla \times (d \v S'/|\v x- \v x'|)= -d \v S' \times\nabla(1/|\v x-\v x'|)$ and $\nabla(1/|\v x- \v x'|)=-\nabla'(1/|\v x- \v x'|)$ in Eq.~(B2), we obtain
\begin{equation}
\oint_{\mathscr{C}}\frac{d \v x'}{|\v x- \v x'|}= \nabla \times \int_{\mathscr{S}}\frac{d\v S'}{|\v x- \v x'|},
\end{equation}
which allows us to write Eq.~(B1) as
\begin{equation}
\v A= \frac{\Phi}{4\pi}\bigg[ \nabla \times \bigg(\nabla \times \int_{\mathscr{S}}\frac{d\v S'}{|\v x- \v x'|}\bigg)\bigg].
\end{equation}
The use of the identity $\nabla^2\v F = \nabla(\nabla \cdot \v F)-\nabla\times (\nabla \times \v F)$ in  Eq.~(B4) gives
\begin{equation}
\v A=\frac{\Phi}{4\pi}\bigg[ \nabla  \int_{\mathscr{S}}\nabla\cdot\bigg(\frac{d\v S'}{|\v x- \v x'|}\bigg)- \int_{\mathscr{S}}\nabla^2\bigg(\frac{d\v S'}{|\v x- \v x'|}\bigg)\bigg].
\end{equation}
Considering the results $\nabla \cdot (d \v S'/|\v x- \v x'|)=\nabla(1/|\v x- \v x'|)\cdot d \v S'$ and $\nabla(1/|\v x- \v x'|)=-\nabla'(1/|\v x- \v x'|)$ in Eq.~(B5), it becomes
\begin{equation}
\v A= \frac{\Phi}{4\pi}\bigg[ -\nabla  \int_{\mathscr{S}}\nabla'\bigg(\frac{1}{|\v x- \v x'|}\bigg) \cdot d\v S'- \int_{\mathscr{S}}\nabla^2\bigg(\frac{d\v S'}{|\v x- \v x'|}\bigg)\bigg].
\end{equation}
Using $\nabla'(1/|\v x- \v x'|)=(\v x- \v x')/|\v x- \v x'|^3$ and $\nabla^2(1/|\v x- \v x'|)=-4\pi \delta(\v x- \v x')$ in Eq.~(B6), we obtain
\begin{equation}
\v A=\frac{\Phi}{4\pi}\bigg[\nabla \int_{\mathscr{S}}\frac{(\v x'-\v x)\cdot d\v S'}{|\v x- \v x'|^3} + 4\pi\int_{\mathscr{S}}\delta(\v x-\v x')\,d\v S'\bigg].
\end{equation}
The first integral is identified with the single-valued solid angle $\Omega_0$ defined in Eq.~(9) while the second integral is identified with the surface vector Dirac delta $\bm{\delta}_{\mathscr{S}}$ specified in Eq.~(10). Thus, we get Eq.~(8):
$\v A = \Phi\nabla\Omega_0/(4\pi) + \Phi \bm{\delta}_{\mathscr{S}}.$

\renewcommand\theequation{C\arabic{equation}}
\setcounter {equation}{0}
\section*{{\large Appendix C. Proof of Eq.~(11)}}
The proof of Eq.~(11) will be developed in two parts. In the first part we will explicitly demonstrate that the circulation of the gradient of $\Omega_0$ along an arbitrary closed path $C$ vanishes $\oint_{C}\nabla\Omega_0\cdot d \v x=0$. In the second part we will transform this circulation using the Stokes theorem $\oint_{C}\nabla\Omega_0\cdot d \v x=0=\int_{S}\nabla \times \nabla\Omega_0 \cdot d \v S$ to show $\nabla \times \nabla \Omega_0=0$. This last result will be used to demonstrate Eq.~(11).

Let us obtain a suitable form of the gradient of $\Omega_0$. The gradient of Eq.~(9) gives
\begin{equation}
\nabla \Omega_0 = \nabla \int_{\mathscr{S}}\frac{(\v x' -\v x)\cdot d \v S'}{|\v x- \v x'|^3}.
\end{equation}
Using $\nabla(1/|\v x- \v x'|)= -(\v x- \v x')/|\v x- \v x'|^3$, Eq.~(C1) becomes
\begin{equation}
\nabla \Omega_0 = -\nabla \int_{\mathscr{S}}\nabla'\bigg(\frac{1}{|\v x- \v x'|}\bigg)\cdot d \v S'.
\end{equation}
Considering the relations $\nabla'(1/|\v x- \v x'|)=-\nabla(1/|\v x- \v x'|)$ and $\nabla(1/|\v x- \v x'|)\cdot d \v S'=\nabla \cdot (d \v S'/|\v x- \v x'|)$ in Eq.~(C2), we obtain
\begin{equation}
\nabla \Omega_0 = \nabla  \int_{\mathscr{S}}\nabla \cdot\bigg(\frac{d \v S'}{|\v x- \v x'|}\bigg).
\end{equation}
When the identity $\nabla(\nabla \cdot \v F) = \nabla \times (\nabla \times \v F) + \nabla^2\v F$ is used in Eq.~(C3), it becomes
\begin{equation}
\nabla \Omega_0 = \nabla \times\bigg[ \nabla \times \int_{\mathscr{S}}\frac{d \v S'}{|\v x- \v x'|}\bigg] + \nabla^2 \int_{\mathscr{S}}\frac{d \v S'}{|\v x- \v x'|}.
\end{equation}
Inserting $\nabla \times (d \v S'/|\v x- \v x'|)= -d \v S' \times\nabla(1/|\v x-\v x'|)$ together with $\nabla(1/|\v x- \v x'|)=-\nabla'(1/|\v x- \v x'|)$ in the quantity within the brackets and using $\nabla^2(1/|\v x- \v x'|)=-4\pi\delta(\v x- \v x')$ on the second term of Eq.~(C4), we obtain
\begin{equation}
\nabla \Omega_0 = \nabla \times\bigg[ \int_{\mathscr{S}}d \v S' \times \nabla'\bigg(\frac{1}{|\v x- \v x'|} \bigg)\bigg] -4\pi \int_{\mathscr{S}}\delta(\v x- \v x')\,d\v S'.
\end{equation}
The quantity within the brackets in Eq.~(C5) can be transformed into a closed line integral via the Stokes theorem
\begin{equation}
\int_{\mathscr{S}}d \v S' \times \nabla'\bigg(\frac{1}{|\v x- \v x'|} \bigg)= \oint_{\mathscr{C}}\frac{d \v x'}{|\v x- \v x'|},
\end{equation}
where $\mathscr{C}$ is the boundary of $\mathscr{S}.$  When Eq.~(C6) and the surface vector Dirac delta $\int_{\mathscr{S}}\delta(\v x- \v x')d\v S'=\bm{\delta}_{\mathscr{S}}$ given in Eq.~(10) are used in Eq.~(C5),  it takes the form
\begin{equation}
\nabla \Omega_0 = \nabla \times\oint_{\mathscr{C}}\frac{d \v x'}{|\v x- \v x'|} -4\pi\bm{\delta}_{\mathscr{S}}.
\end{equation}
Considering $\nabla \times (d \v x'/|\v x - \v x'|)= \nabla(1/|\v x- \v x'|)\times d \v x'$ and $\nabla(1/|\v x- \v x'|)=(\v x'- \v x)/|\v x- \v x'|^3$ we can write Eq.~(C7) as
\begin{equation}
\nabla \Omega_0 = \oint_{\mathscr{C}}\frac{(\v x' - \v x)\times d \v x'}{|\v x- \v x'|^3} -4\pi\bm{\delta}_{\mathscr{S}},
\end{equation}
which is a suitable form of the gradient of $\Omega_0$. Let us now take the circulation to Eq.~(C8) along an arbitrary closed path $C$
\begin{equation}
\oint_{C}\nabla \Omega_0\cdot d \v x = \oint_{C}\oint_{\mathscr{C}}\frac{[(\v x' - \v x)\times d \v x']\cdot d \v x}{|\v x- \v x'|^3} -4\pi\oint_{C}\bm{\delta}_{\mathscr{S}}\cdot d \v x.
\end{equation}
Making use of the relation $[(\v x' - \v x)\times d \v x']\cdot d \v x=(\v x - \v x')\cdot (d\v x \times d\v x')$ in the first term of the right-hand side of Eq.~(C9), we obtain
\begin{equation}
\oint_{C}\nabla \Omega_0\cdot d \v x =4\pi \bigg[\frac{1}{4\pi}\oint_{C}\oint_{\mathscr{C}}\frac{(\v x - \v x')\cdot (d\v x \times d\v x')}{|\v x- \v x'|^3}\bigg] -4\pi\bigg[\oint_{C}\bm{\delta}_{\mathscr{S}}\cdot d \v x\bigg].
\end{equation}
The first quantity within the brackets is the linking number $l$ defined by Eq.~(21). The second quantity within the brackets is another equivalent form of the linking number defined by Eq.~(27). Since $C$ corresponds to the same path in the two closed line integrals on the right-hand side of Eq.~(C10) then it follows that
\begin{equation}
\oint_{C}\nabla \Omega_0 \cdot d\v x=
\begin{cases}
4\pi l - 4\pi l =0& \text{if $C$ encloses $\mathscr{C}$} \\ 0 &\text{otherwise}
\end{cases}
\end{equation}
We observe that regardless of the path $C$ we have the vanishing of the circulation
\begin{equation}
\oint_{C}\nabla \Omega_0 \cdot d\v x=0,
\end{equation}
which is the first step in the proof of Eq.~(11). In the second step we transform the left-hand side of Eq.~(C12) into a surface integral via the Stokes theorem
\begin{equation}
\oint_{C}\nabla \Omega_0 \cdot d\v x= \int_{S} \nabla \times \nabla\Omega_0 \cdot d\v S,
\end{equation}
where $S$ is the surface enclosed by $C.$ Equations (C12) and (C13) imply
\begin{equation}
\oint_{C}\nabla \Omega_0 \cdot d\v x= 0=\int_{S} \nabla \times \nabla\Omega_0 \cdot d\v S.
\end{equation}
Since this result holds for any path $C$ then it follows that the second equality in Eq.~(C14) is valid for any surface $S$ implying the vanishing of the curl of the gradient of $\Omega_0$ in all space
\begin{equation}
\nabla \times \nabla \Omega_0=0.
\end{equation}
To show Eq.~(11) let us write Eq.~(C14) in index notation
\begin{equation}
\oint_{C}\partial_{k}\Omega_0\,dx^k= 0=\int_{S}\varepsilon_{kmn}\partial^m\partial^n\Omega_0\,dS^k.
\end{equation}
Consider now the antisymmetric tensor $dS_{ij}=\varepsilon_{ijk}dS^k$ representing an infinitesimal element of the surface $S$. In terms of $dS_{ij}$ we may write the differential surface vector in the following form $dS^k=(1/2)\varepsilon^{kij}dS_{ij}$. Using this result together with the identity $\varepsilon_{kmn}\varepsilon^{kij}=\delta^{i}_{m}\delta^{j}_{n}-\delta^{j}_{m}\delta^{i}_{n}$ we obtain $\varepsilon_{kmn}\partial^m\partial^n\Omega_0\,dS^k=(1/2)(\partial^i\partial^j - \partial^j \partial^i)\Omega_0 dS_{ij}$, which is used in the second equality in Eq.~(C16) to obtain the relation
\begin{equation}
2\oint_{C}\partial_{k}\Omega_0\,dx^k= 0=\int_{S}\,(\partial^i\partial^j - \partial^j \partial^i)\Omega_0 \,\,dS_{ij}.
\end{equation}
Since the first equality is valid for any path $C$ then the second equality is valid for any surface $S$ and this implies Eq.~(11): $(\partial^i\partial^j - \partial^j \partial^i)\Omega_0=0$ in all space.

\renewcommand\theequation{D\arabic{equation}}
\setcounter {equation}{0}
\section*{{\large Appendix D. Proof of Eq.~(15)}}
Our approach to show Eq.~(15) is as follows. We will show that the circulation of the gradient of the solid angle $\Omega$ along an arbitrary closed path $C$ is non-vanishing: $\oint_{C}\nabla\Omega\cdot d \v x\neq 0$. Then we will transform this circulation via the Stokes theorem $\oint_{C}\nabla\Omega\cdot d \v x=\int_{S}\nabla \times \nabla\Omega \cdot d \v S$ to show $\nabla \times \nabla \Omega= \bm{\delta}_{\mathscr{C}}$. We will use this result to demonstrate Eq.~(15).

Using Eq.~(14), the circulation of the gradient of $\Omega$ takes the form
\begin{equation}
\oint_{C}\nabla\Omega \cdot d \v x= \oint_{C}\nabla\Omega_0 \cdot d \v x + 4\pi\oint_{C}\bm{\delta}_{\mathscr{S}}\cdot d \v x.
\end{equation}
We can transform the first circulation on the right-hand side of Eq.~(D1) using Eq.~(C10). This gives
\begin{eqnarray}
\nonumber \oint_{C}\nabla\Omega \cdot d \v x=\oint_{C}\oint_{\mathscr{C}}\frac{(\v x - \v x')\cdot (d\v x \times d\v x')}{|\v x- \v x'|^3} -4\pi\oint_{C}\bm{\delta}_{\mathscr{S}}\cdot d \v x  + 4\pi\oint_{C}\bm{\delta}_{\mathscr{S}}\cdot d \v x \\
=4\pi \bigg[\frac{1}{4\pi}\oint_{C}\oint_{\mathscr{C}}\frac{(\v x - \v x')\cdot (d\v x \times d\v x')}{|\v x- \v x'|^3}\bigg] .\qquad\qquad\qquad \quad \quad\,\,\,\,\,
\end{eqnarray}
The quantity within the brackets is the Gauss linking number defined in Eq.~(21). Therefore
\begin{equation}
\oint_{C}  \nabla\Omega\cdot d\v x=
\begin{cases}
4\pi l& \text{if $C$ encloses $\mathscr{C}$} \\ 0 &\text{otherwise}
\end{cases}
\end{equation}
Using the Stokes theorem we can transform the left-hand side of Eq.~(D3),
\begin{equation}
\oint_{C}\nabla\Omega \cdot d \v x=\int_{S}\nabla \times \nabla \Omega\cdot d \v S,
\end{equation}
where $C$ is the boundary of the surface $S$. When $C$ does not enclose $\mathscr{C}$ then from Eq.~(D3) we have
\begin{equation}
\oint_{C}\nabla\Omega \cdot d \v x=0=\int_{S}\nabla \times \nabla \Omega\cdot d \v S,
\end{equation}
and thus $\nabla \times \nabla \Omega=0$ locally holds for any surface $S$ not pierced by $\mathscr{C}.$  However, this result does not hold in all space (i.e. globally) because if $C$ encloses $\mathscr{C}$ then the left-hand side of Eq.~(D5) is non-vanishing and from Eq.~(D3) we obtain
\begin{equation}
\oint_{C}\nabla\Omega \cdot d \v x=4\pi l=\int_{S}\nabla \times \nabla \Omega\cdot d \v S,
\end{equation}
and thus the relation $\nabla \times \nabla \Omega \neq 0$ holds. To find the explicit form of this relation, we use Eq.~(14): $\nabla \Omega = \nabla \Omega_0 + 4\pi\bm{\delta}_{\mathscr{S}}$ and therefore $\nabla \times \nabla \Omega = 4\pi\bm{\delta}_{\mathscr{C}}$ where we have used
Eq.~(C15): $\nabla \times \nabla \Omega_0=0$ and Eq.~(12): $\nabla \times  \bm{\delta}_{\mathscr{S}}= \bm{\delta}_{\mathscr{C}}$, where $\bm{\delta}_{\mathscr{C}}=\oint_{\mathscr{C}}\delta(\v x- \v x')d \v x '$ is a line Dirac delta along the closed path $\mathscr{C}$ which forms the boundary of $\mathscr{S}$. Thus
\begin{equation}
\int_{S}\nabla \times \nabla \Omega\cdot d \v S=4\pi\int_{S}\bm{\delta}_{\mathscr{C}}\cdot d \v S,
\end{equation}
which implies
\begin{equation}
\nabla \times \nabla \Omega = 4\pi\bm{\delta}_{\mathscr{C}}.
\end{equation}
To prove Eq.~(15), we write Eq.~(D4) in index notation
\begin{equation}
\oint_{C}\partial_{k}\Omega\,dx^k=\int_{S}\varepsilon_{kmn}\partial^m\partial^n\Omega\,\,dS^k.
\end{equation}
Consider now the antisymmetric tensor $dS_{ij}=\varepsilon_{ijk}dS^k$ representing an infinitesimal element of the surface $S$. Using this result
and the identity $\varepsilon_{kmn}\varepsilon^{kij}=\delta^{i}_{m}\delta^{j}_{n}-\delta^{j}_{m}\delta^{i}_{n}$, we obtain $\varepsilon_{kmn}\partial^m\partial^n\Omega\,dS^k=(1/2)(\partial^i\partial^j - \partial^j \partial^i)\Omega dS_{ij}$, which is used in the second equality in Eq.~(D9), obtaining
\begin{equation}
2\oint_{C}\partial_{k}\Omega \,dx^k=\int_{S}(\partial^i\partial^j - \partial^j \partial^i)\Omega \,\,dS_{ij}.
\end{equation}
When the path $C$ does not enclose the curve $\mathscr{C}$ then from Eq.~(D3) we have
\begin{equation}
2\oint_{C}\partial_{k}\Omega \,dx^k=0=\int_{S}(\partial^i\partial^j - \partial^j \partial^i)\Omega \,\,dS_{ij},
\end{equation}
which implies $(\partial^i\partial^j - \partial^j \partial^i)\Omega=0$ for any surface $S$ not pierced by $\mathscr{C}.$ In this case $\Omega$ is locally single-valued. However, this is not the global case for if $C$ encloses $\mathscr{C}$ then the left-hand side of Eq.~(D10) is non-vanishing and from Eq.~(D3) we obtain
\begin{equation}
2\oint_{C}\partial_{k}\Omega\, dx^k=8\pi l=\int_{S}(\partial^i\partial^j - \partial^j \partial^i)\Omega \,\,dS_{ij}.
\end{equation}
which implies $(\partial^i\partial^j - \partial^j \partial^i)\Omega\neq 0$ when $C$ encircles $\mathscr{C},$ or equivalently stated, when $C$ crosses $S$. To find the explicit form of  $(\partial^i\partial^j - \partial^j \partial^i)\Omega$ we use Eqs.~(D12) and (D9) to obtain
\begin{equation}
\int_{S}(\partial^i\partial^j - \partial^j \partial^i)\Omega \,\,dS_{ij}=2\int_{S}\varepsilon_{kmn}\partial^m\partial^n\Omega\,\,dS^k.
\end{equation}
Equation~(D8) in index notation reads $4\pi(\bm{\delta}_{\mathscr{C}})_k=\varepsilon_{kmn}\partial^m\partial^n\Omega$. This result and  $dS^k=(1/2)\varepsilon^{kij}dS_{ij}$ give the relation $\varepsilon_{kmn}\partial^m\partial^n\Omega\,dS^k=2\pi\varepsilon^{ijk}(\bm{\delta}_{\mathscr{C}})_kdS_{ij}$ so that Eq.~(D13) reduces to
\begin{equation}
\int_{S}\,(\partial^i\partial^j - \partial^j \partial^i)\Omega \,\,dS_{ij}= 4\pi\int_{S} \varepsilon^{ijk}(\bm{\delta}_{\mathscr{C}})_k \,dS_{ij},
\end{equation}
and this implies Eq.~(15): $(\partial^i\partial^j - \partial^j \partial^i)\Omega=4\pi\varepsilon^{ijk}(\bm{\delta}_{\mathscr{C}})_k.$

\renewcommand\theequation{E\arabic{equation}}
\setcounter {equation}{0}
\section*{{\large Appendix E. Proofs of Eqs.~(74), (77), and (80)}}
The proof of Eq.~(74) is based on a proof given by Gelca \cite{41}. Similar proofs for Eqs.~(77) and (80) will be given. Our general strategy is as follows: we will apply topological transformations to the linking number $l$ (i.e. deformations of the associated curves in $l$) and show that these transformations leave the linking number invariant.
\vskip 4pt
\noindent {\bf Proof of Eq.~(74).} Consider the linking number of the curves ${\mathbb C}$ and $\mathscr{C}$
\begin{equation}
\frac{1}{4\pi}\oint_{{\mathbb C}}\oint_{\mathscr{C}}\frac{(\v x - \v x')\cdot (d\v x \times d\v x')}{|\v x- \v x'|^3}=l({\mathbb C},\mathscr{C}).
\end{equation}
Let $C$ be a closed path encircling the curve $\mathscr{C}$ and let us deform the path $C$ into the path $C'$ via the transformation $C \to C'$ and let ${\mathbb C}=C\cup(-C')$ be the union of $C$ and $(-C')$ which bounds the surface ${\mathbb S}$ traced by $C$ while being deformed into $C'$. Accordingly, ${\mathbb C}=C\cup(-C')=\partial {\mathbb S}$ where $\partial {\mathbb S}$ is the boundary of ${\mathbb S}$. We also assume $C$ and $C'$ encircle the same number of times $\mathscr{C}$. Using the properties $\oint_{{\mathbb C}\,=\,C\,\cup\,(-C')} = \oint_{C} + \oint_{-C'}$ and $\oint_{-C'} = -\oint_{C'}$ it follows that Eq.~(E1) can be decomposed as
\begin{eqnarray}
\nonumber \frac{1}{4\pi}\oint_{{\mathbb C}}\oint_{\mathscr{C}}\frac{(\v x - \v x')\cdot (d\v x \times d\v x')}{|\v x- \v x'|^3}=\frac{1}{4\pi}\oint_{C}\oint_{\mathscr{C}}\frac{(\v x - \v x')\cdot (d\v x \times d\v x')}{|\v x- \v x'|^3}\quad \\
- \frac{1}{4\pi}\oint_{C'}\oint_{\mathscr{C}}\frac{(\v x - \v x')\cdot (d\v x \times d\v x')}{|\v x- \v x'|^3},
\end{eqnarray}
or equivalently,
\begin{equation}
l({\mathbb C},\mathscr{C})= l(C,\mathscr{C}) - l(C',\mathscr{C}),
\end{equation}
where
\begin{eqnarray}
\nonumber \frac{1}{4\pi}\oint_{C}\oint_{\mathscr{C}}\frac{(\v x - \v x')\cdot (d\v x \times d\v x')}{|\v x- \v x'|^3}=l(C,\mathscr{C}),\\ \frac{1}{4\pi}\oint_{C'}\oint_{\mathscr{C}}\frac{(\v x - \v x')\cdot (d\v x \times d\v x')}{|\v x- \v x'|^3}=l(C',\mathscr{C}),
\end{eqnarray}
are the linking numbers of $C$ and $\mathscr{C}$, and $C'$ and $\mathscr{C}$, respectively. Therefore if $l({\mathbb C},\mathscr{C})=0$ then $l(C,\mathscr{C})=l(C',\mathscr{C})$ and this would proof Eq.~(74). In Appendix C we demonstrated Eq.~(C8) which can be re-arranged to obtain the relation
\begin{equation}
\oint_{\mathscr{C}}\frac{(\v x' - \v x)\times d \v x'}{|\v x- \v x'|^3}= \nabla \Omega_0(\mathscr{S}) + 4\pi \bm{\delta}_{\mathscr{S}},
\end{equation}
where $\Omega_0(\mathscr{S}) = \int_{\mathscr{S}}\{(\v x' - \v x)\cdot d \v S'/|\v x- \v x'|^3 \}$ is the single-valued solid angle subtended by $\mathscr{C}$ and $\bm{\delta}_{\mathscr{S}}=\int_{\mathscr{S}}\delta(\v x- \v x')d \v S'$ is the surface vector Dirac delta defined along the surface $\mathscr{S}$ bounded by $\mathscr{C}.$ Using Eq.~(E5) in the left-hand side of Eq.~(E1) and $[(\v x' - \v x)\times d \v x']\cdot d \v x=(\v x - \v x')\cdot (d\v x \times d\v x')$ it follows
\begin{equation}
l({\mathbb C},\mathscr{C})= \frac{1}{4\pi}\oint_{{\mathbb C}}\nabla \Omega_0(\mathscr{S})\cdot d\v x +  \oint_{{\mathbb C}}\bm{\delta}_{\mathscr{S}}\cdot d \v x.
\end{equation}
The first line integral in the right-hand side vanishes because $\Omega_0(\mathscr{C})$ is a single-valued function. Applying the Stokes theorem to the second line integral in the right-hand side, we obtain
\begin{equation}
l({\mathbb C},\mathscr{C})= \int_{{\mathbb S}}\bm{\delta}_{\mathscr{C}}\cdot d \v S,
\end{equation}
where ${\mathbb S}$ is the surface bounded by ${\mathbb C}$ and we have used Eq.~(12): $\nabla \times \bm{\delta}_{\mathscr{S}} = \bm{\delta}_{\mathscr{C}}$ where $\bm{\delta}_{\mathscr{C}}= \oint_{\mathscr{C}}\delta(\v x- \v x')d \v x'$ is a line vector Dirac delta defined along $\mathscr{C}.$ The surface ${\mathbb S}$ corresponds to the surface traced by the path $C$ while being deformed into the path $C'$ and therefore the curve $\mathscr{C}$ never crosses the surface ${\mathbb S}$. Accordingly, the function $\bm{\delta}_{\mathscr{C}}$ vanishes along the surface ${\mathbb S}$ and therefore $\int_{{\mathbb S}}\bm{\delta}_{\mathscr{C}}\cdot d \v S=0$ which gives $l({\mathbb C},\mathscr{C})=0$. This result and Eq.~(E3) imply $l(C,\mathscr{C})=l(C',\mathscr{C})$ and this proves Eq.~(74).
\vskip 4pt
\noindent {\bf Proof of Eq.~(77).} Consider the linking number of the curves $C$ and ${\cal C}$
\begin{equation}
\frac{1}{4\pi}\oint_{C}\oint_{{\cal C}}\frac{(\v x - \v x')\cdot (d\v x \times d\v x')}{|\v x- \v x'|^3}=l(C,{\cal C}).
\end{equation}
Let $C$ be a closed path encircling the curve $\mathscr{C}$. Let us deform the curve $\mathscr{C}$ into the curve $\mathscr{C}'$ via the transformation $\mathscr{C} \to \mathscr{C}'$ and let ${\cal C}=\mathscr{C}\cup(-\mathscr{C'})$ be the union of $\mathscr{C}$ and $(-\mathscr{C'})$ which bounds the surface ${\cal S}$ traced by $\mathscr{C}$ while being deformed into $\mathscr{C}'$. It follows that ${\cal C}=\mathscr{C}\cup(-\mathscr{C'})=\partial {\cal S}$ where $\partial {\cal S}$ is the boundary of ${\cal S}$. We assume $C$ encircles the same number of times $\mathscr{C}$ and $\mathscr{C}'$. Using the properties $\oint_{{\cal C}\,=\,\mathscr{C}\,\cup\,(-\mathscr{C}')} = \oint_{\mathscr{C}} + \oint_{-\mathscr{C}'}$ it follows that Eq.~(E8) can be decomposed as
\begin{eqnarray}
\nonumber \frac{1}{4\pi}\oint_{C}\oint_{{\cal C}}\frac{(\v x - \v x')\cdot (d\v x \times d\v x')}{|\v x- \v x'|^3}=\frac{1}{4\pi}\oint_{C}\oint_{\mathscr{C}}\frac{(\v x - \v x')\cdot (d\v x \times d\v x')}{|\v x- \v x'|^3}\quad \\
- \frac{1}{4\pi}\oint_{C}\oint_{\mathscr{C}'}\frac{(\v x - \v x')\cdot (d\v x \times d\v x')}{|\v x- \v x'|^3},
\end{eqnarray}
or equivalently,
\begin{equation}
l(C,{\cal C})= l(C,{\mathscr{C}})- l(C,\mathscr{C}'),
\end{equation}
where
\begin{eqnarray}
\nonumber \frac{1}{4\pi}\oint_{C}\oint_{\mathscr{C}}\frac{(\v x - \v x')\cdot (d\v x \times d\v x')}{|\v x- \v x'|^3}=l(C,\mathscr{C}), \\\frac{1}{4\pi}\oint_{C}\oint_{\mathscr{C}'}\frac{(\v x - \v x')\cdot (d\v x \times d\v x')}{|\v x- \v x'|^3}=l(C,\mathscr{C}'),
\end{eqnarray}
are the linking numbers of $C$ and $\mathscr{C}$, and $C$ and $\mathscr{C}'$, respectively.  Therefore if $l(C, {\cal C})=0$ then $l(C,\mathscr{C})=l(C,\mathscr{C}')$ and this would prove Eq.~(77). Following the same line of arguments that led to Eq.~(C8), it follows that we can make the replacement $\mathscr{S}\to\cal{S}$ in Eq.~(C8) and obtain
\begin{equation}
\oint_{\cal{C}}\frac{(\v x' - \v x)\times d \v x'}{|\v x- \v x'|^3}= \nabla \Omega_0({\cal{S}}) + 4\pi \bm{\delta}_{\cal{S}},
\end{equation}
where $\Omega_0({\cal{S}}) = \int_{\cal{S}}\{(\v x' - \v x)\cdot d \v S'/|\v x- \v x'|^3 \}$ is the single-valued solid angle function subtended by the curve $\cal{C}$ and $\bm{\delta}_{\cal{S}}=\int_{\cal{S}}\delta(\v x- \v x')d \v S'$ is the surface vector Dirac delta defined along the surface $\cal{S}$ bounded by $\cal{C}.$ Using Eq.~(E12) and the relation  $[(\v x' - \v x)\times d \v x']\cdot d \v x=(\v x - \v x')\cdot (d\v x \times d\v x')$ we obtain
\begin{equation}
l(C,{\cal{C}})= \frac{1}{4\pi}\oint_{C}\nabla \Omega_0({\cal{S}})\cdot d \v x + \oint_{C}\bm{\delta}_{\cal{S}}\cdot d \v x.
\end{equation}
The first line integral on the right-hand side vanishes because $\Omega_0(\cal{S})$ is single-valued. On the other hand, the surface $\cal S$ corresponds to the surface traced by the curve $\mathscr{C}$ while being deformed into the curve $\mathscr{C}'$ and therefore the path $C$ never crosses the surface $\cal{S}$. Consequently, the function $\bm{\delta}_{\cal S}$ vanishes along the path $C$ so that $\oint_{C}\bm{\delta}_{\cal{S}}\cdot d \v x=0$ and this gives $l(C,{\cal{C}})=0.$ This result and the right-hand side of Eq.~(E10) imply $l(C,{\mathscr{C}})=l(C,\mathscr{C}')$, result that  proves Eq.~(77).
\vskip 4pt
\noindent {\bf Proof of Eq.~(80).} Consider the linking number of the curves ${\mathbb{C}}$ and $\cal{C}$
\begin{equation}
\frac{1}{4\pi}\oint_{{\mathbb C}}\oint_{\mathscr{C}}\frac{(\v x - \v x')\cdot (d\v x \times d\v x')}{|\v x- \v x'|^3}=l({\mathbb C},\cal{C}).
\end{equation}
Let $C$ be a closed path encircling the curve $\mathscr{C}$. Let us simultaneously deform the path $C$ into the path $C'$ via the transformation $C \to C'$ and deform the curve $\mathscr{C}$ into the curve $\mathscr{C}'$ via the transformation $\mathscr{C}\to\mathscr{C}'.$ Let ${\mathbb C}=C\cup(-C')$ be the union of $C$ and $(-C')$ which bounds the surface ${\mathbb S}$ traced by $C$ while being deformed into $C'$ and let ${\cal C}=\mathscr{C}\cup(-\mathscr{C'})$ be the union of $\mathscr{C}$ and $(-\mathscr{C'})$ which bounds the surface ${\cal S}$ traced by $\mathscr{C}$ while being deformed into $\mathscr{C}'$. Accordingly, ${\mathbb C}=C\cup(-C')=\partial {\mathbb S}$ where $\partial {\mathbb S}$ is the boundary of ${\mathbb S}$ and ${\cal C}=\mathscr{C}\cup(-\mathscr{C'})=\partial {\cal S}$ where $\partial {\cal S}$ is the boundary of ${\cal S}$. We assume the path $C$ encircles $\mathscr{C}$ and $\mathscr{C}'$ the same number of times the path $C'$ encircles $\mathscr{C}$ and $\mathscr{C}'$. Using  $\oint_{{\mathbb C}\,=\,C\,\cup\,(-C')}\oint_{\cal{C}\,=\,\mathscr{C}\,\cup\,(-\mathscr{C}')}=(\oint_{C} - \oint_{C'})(\oint_{\mathscr{C}} - \oint_{\mathscr{C}'})=\oint_{C}\oint_{\mathscr{C}} - \oint_{C}\oint_{\mathscr{C}'} -\oint_{C'}\oint_{\mathscr{C}} +  \oint_{C'}\oint_{\mathscr{C}'}$, $\oint_{{\mathbb C}\,=\,C\,\cup\,(-C')}=\oint_{C} + \oint_{-C'}$, $\oint_{-C'}=-\oint_{C'}$, $\oint_{{\cal{C}}\,=\,\mathscr{C}\,\cup\,(-\mathscr{C}')}=\oint_{\mathscr{C}} + \oint_{-\mathscr{C}}$, and $\oint_{-\mathscr{C}'}=-\oint_{\mathscr{C}'}$, we can decompose Eq.~(E14) as
\begin{eqnarray}
\nonumber \frac{1}{4\pi}\oint_{{\mathbb C}}\oint_{\cal{C}}\frac{(\v x - \v x')\cdot (d\v x \times d\v x')}{|\v x- \v x'|^3}\qquad\qquad\qquad\qquad\qquad\qquad\qquad\qquad\qquad\qquad\\
\nonumber = \frac{1}{4\pi}\oint_{C}\oint_{\mathscr{C}}\frac{(\v x - \v x')\cdot (d\v x \times d\v x')}{|\v x- \v x'|^3} -\frac{1}{4\pi} \oint_{C}\oint_{\mathscr{C}'}\frac{(\v x - \v x')\cdot (d\v x \times d\v x')}{|\v x- \v x'|^3}\qquad\quad\\
 -\frac{1}{4\pi}\!\oint_{C'}\oint_{\mathscr{C}}\frac{(\v x - \v x')\cdot (d\v x \times \!d\v x')}{|\v x- \v x'|^3} + \frac{1}{4\pi}\oint_{C'}\oint_{\mathscr{C}'}\frac{(\v x - \v x')\cdot (d\v x \times d\v x')}{|\v x- \v x'|^3}, \qquad
\end{eqnarray}
or equivalently,
\begin{equation}
l({\mathbb{C}},{\cal C})= l(C,{\mathscr{C}})-l(C,{\mathscr{C}'})-l(C',{\mathscr{C}})+l(C',{\mathscr{C}'}),
\end{equation}
where the corresponding linking numbers are defined by
\begin{eqnarray}
\nonumber \frac{1}{4\pi}\oint_{C}\oint_{\mathscr{C}}\frac{(\v x - \v x')\cdot (d\v x \times d\v x')}{|\v x- \v x'|^3}=l(C,\mathscr{C}),\\ \frac{1}{4\pi} \oint_{C}\oint_{\mathscr{C}'}\frac{(\v x - \v x')\cdot (d\v x \times d\v x')}{|\v x- \v x'|^3}=l(C,\mathscr{C}'),\\
 \nonumber \frac{1}{4\pi}\oint_{C'}\oint_{\mathscr{C}}\frac{(\v x - \v x')\cdot (d\v x \times d\v x')}{|\v x- \v x'|^3}=l(C',\mathscr{C}), \\ \frac{1}{4\pi}\oint_{C'}\oint_{\mathscr{C}'}\frac{(\v x - \v x')\cdot (d\v x \times d\v x')}{|\v x- \v x'|^3}=l(C',\mathscr{C}').
\end{eqnarray}
Now, we have the result $l(C',\mathscr{C})=l(C,\mathscr{C})$ because of Eqs.~(E3) and (E7) (which follows from the transformation $C\to C'$). Also, we have the result $=l(C,{\mathscr{C}})=l(C,{\mathscr{C}'})$ because of Eq.~(E10) and (E13) (which follows from the transformation $\mathscr{C}\to\mathscr{C}'$). Using these results Eq.~(E16) reduces to
\begin{equation}
l({\mathbb{C}},{\cal C})= l(C',{\mathscr{C}'})-l(C,{\mathscr{C}}).
\end{equation}
Accordingly, if the left-hand side of Eq.~(E19) vanishes then $l(C,{\mathscr{C}})=l(C',{\mathscr{C}'})$ and this would prove  Eq.~(80). Using Eq.~(E12) together with $[(\v x' - \v x)\times d \v x']\cdot d \v x=(\v x - \v x')\cdot (d\v x \times d\v x')$ we can write
\begin{equation}
l({\mathbb{C}},{\cal C})= \frac{1}{4\pi}\oint_{{\mathbb C}}\nabla \Omega_0({\cal{S}})\cdot d \v x + \oint_{{\mathbb C}}\bm{\delta}_{\cal{S}}\cdot d \v x.
\end{equation}
The first line integral on the right-hand side vanishes because $\Omega_0(\cal{S})$ is single-valued. This result and the relations $\oint_{{\mathbb C}\,=\,C\,\cup\,(-C')}=\oint_{C} + \oint_{-C'}$ and $\oint_{-C'}=-\oint_{C'}$ yield $l({\mathbb{C}},{\cal C})=\oint_{C}\bm{\delta}_{\cal{S}}\cdot d \v x-\oint_{C'}\bm{\delta}_{\cal{S}}\cdot d \v x.$ The surface $\cal{S}$ corresponds to the surface traced by the curve $\mathscr{C}$ while being deformed into the curve $\mathscr{C}'.$ Accordingly, neither the path $C$ nor the path $C'$ cross the surface $\cal{S}$ along which $\bm{\delta}_{\cal{S}}$ is non-vanishing. Therefore $\oint_{C}\bm{\delta}_{\cal{S}}\cdot d \v x=0$ and $\oint_{C'}\bm{\delta}_{\cal{S}}\cdot d \v x=0$ which implies $l({\mathbb{C}}, {\cal C})=0.$ This result and Eq.~(E19) give $l(C,{\mathscr{C}})=l(C',{\mathscr{C}'})$ and this proves Eq.~(80).

\end{document}